\documentclass[aps,amsfonts]{revtex4}
\usepackage{graphics}
\usepackage{graphicx}

\begin{document}

\title{Improved gravitational waveforms from spinning black hole binaries}
\author{Edward K. Porter$^{1,2}$ and B. S. Sathyaprakash$^{1}$} 
\affiliation{$^{1}$School of Physics and Astronomy, Cardiff University, 5, The 
Parade, Cardiff, Wales, UK, CF24 3YB}
\affiliation{$^{2}$Laboratoire de l'Acc\'el\'erateur Lin\'eaire, B.P. 34, 
B\^{a}timent 208, Campus d'Orsay, 91898 Orsay Cedex, France}
\vspace{1cm}
\begin{abstract}
\noindent The standard post-Newtonian approximation to gravitational waveforms,
called T-approximants, from non-spinning black hole binaries are known not 
to be sufficiently accurate close to the last stable orbit of the system.  
A new approximation, called P-approximants,  is believed to improve the 
accuracy of the waveforms rendering them applicable up to the last stable 
orbit.  In this study we apply P-approximants to the case of a 
test-particle in equatorial orbit around a Kerr black hole parameterized
by a spin parameter $q$ that takes values between $-1$ and $1.$
In order to assess the performance of the two approximants we 
measure their {\it effectualness} (i.e. larger overlaps with the exact signal), 
and {\it faithfulness} (i.e. smaller biases while measuring the parameters of
the signal) with the exact (numerical) waveforms.  We find that in the 
case of prograde orbits, that is orbits whose angular momentum is in 
the same sense as the spin angular momentum of the black hole, 
T-approximant templates obtain an effectualness of $\sim 0.99$
for spins $q \lesssim 0.75.$ For $0.75 < q < 0.95,$ the 
effectualness drops to about 0.82.  The P-approximants achieve effectualness
of $> 0.99$ for all spins up to $q = 0.95.$  The 
bias in the estimation of parameters is much lower in the case of P-approximants
than T-approximants.  We find that P-approximants are both effectual 
and faithful and should be more effective than T-approximants
as a detection template family when $q>0.$
For $q<0$ both T- and P-approximants perform equally well
so that either of them could be used as a detection template family.
\end{abstract}

\maketitle

\section{Introduction}
Stellar mass compact binaries consisting of double neutron stars (NS),
double black holes (BH) or a mixed binary consisting of a neutron star 
and a black hole, are  the primary targets for a direct first detection of 
gravitational waves (GW) by interferometric detectors, LIGO \cite{LIGO}, VIRGO
\cite{VIRGO},  GEO600 \cite{GEO}, and TAMA~\cite{TAMA}. 
Under radiation reaction the orbit of a binary slowly decays, emitting a signal 
whose amplitude and frequency 
increases with time and is termed a ``chirp'' signal. 
While it is believed that there is a greater population of NS-NS binaries 
\cite{Grish,Phin,Narayan,Stairs,KalogeraAndBelczynski},  
it is the BH-BH binaries that are the strongest candidates for detection since
they can be seen from a greater volume, about two orders-of-magnitude greater
than NS-NS binaries \cite{PostnovEtAl,Grish}.

\subsection{Matched filtering}
In order to detect such sources one employs the method of matched 
filtering ~\cite{Helst}.  Briefly, the method works as follows: 
Firstly, one creates a set of waveforms, or templates
as they are called, that depend on a number of parameters of the
source and its location and orientation relative to the detector. 
These templates are then cross-correlated with 
the detector output weighted by the inverse of the noise spectral density.  
If a signal, whose parameters are close to one of the template waveforms,
is actually present in the detector output then the cross-correlation builds
up, with the dominant contribution coming from frequencies where the noise
spectral density is low.  Thus, in the presence of a sufficiently strong
signal the correlation will be much larger than the RMS correlation in the
absence of any signal.
How large should it be before we can be confident about the presence of
a signal depends on the combination of the rate of inspiral events and
the false alarm probability (see e.g. Ref. \cite{DhurSath} for a simple estimation).
 
The effectiveness of matched filtering depends on how well the phase 
evolution of the waveform is known. Even tiny instantaneous differences,
as low as one part in $10^3$ in the phase of the true signal that
might be present in the detector output and the template that is 
used to dig it out could lead to a cumulative difference of several
radians since one integrates over several hundreds to several thousands
of cycles. In view of improving the signal-to-noise ratio for inspiral
events there has been a world-wide effort in accurately computing the 
dynamics of a compact binary and the waveform it emits or to use
phenomenologically defined detection template families \cite{BCV1,BCV2,BCV3}. 

\subsection{Phasing of the coalescing binary signal}

There have been parallel efforts on using two different approximation schemes: On the
one hand the post-Newtonian (PN) expansion of Einstein's equations has been used
to treat the dynamics of two bodies of comparable masses with and
without spin, in orbit around
each other. This approximation is applicable when the velocities involved 
in the system are small but there is no restriction on the ratio of the
masses \cite{BDIWW,BDI,WillWise,BIWW,Blan1,DJSABF,BFIJ}.  

On the other hand, black hole perturbation theory has been used to compute
the dynamics of a test particle in orbit around a spin-less or spinning
black hole. Black hole perturbation theory does not make any assumptions 
on the velocity of the components, but is valid only in the limit when
the mass of one of the bodies is much less than the other
\cite{Poisson1,Cutetal1,TagNak,Sasaki,TagSas,TTS}.  

The post-Newtonian approximation is a perturbative method which expands the equations of 
motion, binding energy and GW flux as a power series in $v/c$, where $v$ is a 
typical velocity in the system and $c$ is the speed of light.  In the early stages of an 
inspiral, the radiation reaction time-scale $\tau_{\rm RR} \sim \omega/\dot\omega,$ where
$\omega$ is the angular velocity and $\dot \omega$ its time-derivative, is much greater than 
the orbital time-scale $\tau_{\rm orb} \sim 1/\omega$.  It is during this adiabatic regime that 
the post-Newtonian approximation works best.  
At present, the PN expansion for the case of comparable-masses is known to order 
${\mathcal O}\left(v^{6}\right)$~\cite{DJSABF} and ${\mathcal 
O}\left(v^{7}\right)$~\cite{BFIJ}, for the energy and flux functions,
respectively.  However, at this order an arbitrary parameter exists 
in the expression for the flux.  In order to see how well PN theory performs, we can 
compare two different systems.  If we assume a NS-NS binary of masses 
(1.4,\,1.4)~$M_{\odot}$ and a lower frequency cutoff of the detector at 40 Hz, then the 
``orbital velocity" of the binary is small, $v \sim 0.12$, \footnote{We shall work
in a system of units in which the speed of light and Newton's gravitational 
constant are both set equal to unity: $c=G=1.$}  when it enters 
the detector bandwidth and the two stars are still largely separated,
$r \sim 70\,M$.  The ratio of time-scales in the most sensitive 
regime of the detector is in the range
$4.5\times10^{3}\leq\tau_{\rm RR}/\tau_{\rm orb}\leq 680$.  If on the other hand we 
take a BH-BH binary of masses (10,\,10)\,$M_{\odot}$, the orbital velocity is
quite large,  $v \sim 0.23,$ and the separation is quite small, $r \sim 19\,M,$ 
upon entering the detector bandwidth.  This is very close to the regime, 
$v \sim 0.3$, $r \sim 11 M$, where the background curvature becomes strong 
and the motion relativistic. Once again, comparing time-scales, we obtain 
$140\leq\tau_{\rm RR}/\tau_{\rm orb}\leq 40$, where the final value is taken at the 
last stable orbit  at $f_{\rm LSO} \sim 210$ Hz.  It is known that PN theory becomes 
inaccurate at an orbital separation of $r \leq 10\,M$~\cite{Brady}.  
Therefore, post-Newtonian approximation becomes less valid for higher mass systems in the LIGO band but well describes the 
early stages of the inspiral of a NS-NS system visible in LIGO.

As previously stated, black hole perturbation theory makes no assumptions about the orbital 
velocity of the components, but does restrict their masses.  One assumes that 
a test particle of mass $\mu$ is in orbit about a central BH of mass $M$ 
such that $\mu \ll M$.  Assuming this restriction is satisfied we have an analytical 
expression for the energy.  However, no analytical expression has been worked
out for the gravitational wave flux emitted by such a system.  Using black 
hole perturbation theory, a series approximation was initially calculated 
to ${\mathcal O}\left(v^{3}\right)$ by Poisson for a test particle in circular 
orbit around a Schwarzschild black hole~\cite{Poisson1}.  
The series was extended numerically to ${\mathcal O}\left(v^{5}\right)$ by 
Cutler et al.~\cite{Cutetal1}, and then to ${\mathcal O}\left(v^{8}\right)$ by 
Tagoshi and Nakamura~\cite{TagNak}  and confirmed analytically 
by Tagoshi and Sasaki~\cite{TagSas}.  The most recent 
progress is an extension of the series to ${\mathcal O}\left(v^{11}\right)$ by 
Tagoshi, Tanaka and Sasaki~\cite{TTS}.  For a test particle in circular orbit 
about a Kerr black hole, the initial progress was again made by 
Poisson~\cite{Poisson2}.  The series approximation was improved from ${\mathcal 
O}\left(v^{3}\right)$ to ${\mathcal O}\left(v^{5}\right)$, and subsequently to 
${\mathcal O}\left(v^{8}\right),$ by Tagoshi, Tanaka, Shibata and 
Sasaki~\cite{SSTT,TSTS}.  

Several authors \cite{TTS,Cutetal2,Poisson3,Poisson4,DIS1} 
have shown that the convergence of both post-Newtonian 
approximation and black hole perturbation theory is too slow 
to be useful in constructing accurate templates.  
More recently,  Damour, Iyer and Sathyaprakash 
(hereafter DIS) showed for the case of a test-mass in orbit about a 
Schwarzschild BH, that by using properly defined energy and flux functions
that have better analytical properties, combined with Pad\'e techniques, it was 
possible to take the existing series expansion and improve its convergence 
properties~\cite{DIS1}.  The new approximation in which Pad\'e approximants
of new energy and flux functions are used to derive improved templates is called
P-approximant.  Using a fiducially defined ``exact'' waveform, it was shown 
that the P-approximant templates were both more {\it effectual} 
(i.e. larger overlaps with the exact waveform) and {\it faithful} 
(i.e. larger overlaps with the exact waveform and lower biases 
in the estimation of parameters) than the corresponding post-Newtonian  
(hereafter T-approximant) templates.  While in general, more templates are 
needed for P-approximant templates to cover the same volume of parameter 
space~\cite{EKP}, the extra computational cost is preferred
for the increased performance in P-approximants.  
The failure of the PN 
expansion to converge sufficiently quickly in the case of a test particle 
orbiting a Schwarzschild BH~\cite{BHChapter} does not bode well for the 
modelling of even a test particle inspiralling into a Kerr BH.  Another 
motivation for this work is that at present there is effort to use
{\it effective one body} (EOB) waveforms~\cite{BD,Damour01} to detect the inspiral and merger
signals from two comparable-mass Kerr BHs in GW data.  As the EOB waveforms are based 
partially on P-approximants, any development and concretization of the benefits of 
P-approximant templates will boost our confidence in using these improved waveforms
as detection templates.

\subsection{Organisation of the paper}

In this paper we will extend the P-approximant technique to the case of 
a test particle orbiting a Kerr black hole.  The reason for focusing on 
test-mass systems is that we can use the exact numerical fluxes \cite{Shibata} from black 
hole perturbation theory with which to compare our results and thereby
reliably demonstrate the usefulness of the technique.  We begin in 
Sec.~\ref{sec:waveform} with a summary of the
matched filtering and the signal-to-noise ratio achieved by the first generation
of GEO, LIGO and VIRGO interferometers for spinning black hole binaries.

We then go on to discuss in Sec.~\ref{sec:EnergyAndFlux}
the current state-of-the-art in our understanding of
the evolution of a test particle in orbit around a Kerr black hole. In particular, 
we shall discuss the time-evolution of the orbital energy and gravitational 
wave flux as a function of the spin of the central black hole at various
post-Newtonian orders, and the locations of the last stable and unstable circular 
orbits.  We shall see that the post-Newtonian expansion of the
flux does not show a regular behaviour as we move from low to high orders
in the post-Newtonian expansion, becoming worse for more rapidly spinning
black holes.  
In order to improve the convergence properties of the flux function,
in Sec.~\ref{sec:fluxmodel} we shall introduce
a modified form of the flux function and its Pad\'e approximant. We shall
demonstrate in Sec.~\ref{sec:results} the improved behaviour of the Pad\'e approximant,
 at first graphically and then by showing that the overlaps, of the inspiral waveform 
based on it with the exact waveform, are close to unity.
We shall use a number of different test systems in our comparison:  
These will range from systems with dissimilar masses, such as a 
NS-BH binary, to comparable-mass systems, such as a BH-BH binary. 
To deal with the comparable mass systems, we shall introduce in
the energy and flux functions a term dependent on the symmetric mass ratio 
$\eta \equiv m_1m_2/M^2.$ While not being entirely consistent, because of the fact 
that no finite mass correction terms are included, it allows us to examine the 
performance of various templates as we move from the test-mass systems to 
comparable-mass systems.  

The emphasis of the current paper is also on the estimation of parameters. We have
carried out a detailed study of how good T- and P-approximants are in measuring 
the parameters of the binary.  We shall show in Sec.~\ref{sec:results} 
that as a result of not being {\it effectual} representations
of the exact signal, T-approximants also turn out not to be  {\it faithful} representations
either. In other words, the systematic error in the estimation of the parameters 
caused by the wrong phasing of the signal is much larger in the case of T-approximants 
than in the case of P-approximants. In summary, P-approximants are not only {\it effectual}
but they are {\it faithful} too as in the case of non-spinning black hole binaries.

\section{Waveform and Signal-to-Noise Ratio} \label{sec:waveform}
In this Section we will discuss the nature of the post-Newtonian waveform
from an inpiralling compact binary and the response of an antenna to such
a signal. We will then use the Fourier transform of the waveform to compute
the signal-to-noise ratio (SNR) that can be achieved for these signals
when they are detected using matched filtering.

\subsection{The Waveform} 
\label{sec:the waveform}
In the transverse-traceless gauge gravitational waves are represented 
by two polarisation amplitudes $h_+$ and $h_\times.$ 
The response $h(t)$ of an antenna to an incoming signal is expressed 
as a combination of the two polarisation states and the beam pattern 
functions $F_+$ and $F_\times$ of the antenna as \cite{Thorne}: 
$h(t) = h_+ F_+ + F_\times h_\times.$ For a wave from a binary of 
masses $m_1$ and $m_2$ (total mass $M=m_1+m_2$ and
mass ratio $\eta = m_1 m_2/M^2$) that is inclined with respect to the 
plane of the sky at an angle $i$, propagating in the direction 
$(\theta,\, \bar{\phi}),$ (see Ref.~\cite{Thorne} for exact definitions),
frequency $F$ and polarization angle $\bar{\psi}$ with respect to the antenna, 
the polarisation amplitudes, in the so-called {\it restricted} post-Newtonian 
approximation \cite{Cutetal2}, are
\begin{eqnarray}
h_+ (t;\, M,\eta,i) & = & \frac{4\eta M}{d} \frac{(1+\cos^2 i)}{2}\, v_F^2(t)\, \cos \phi(t), \\
h_\times(t;\, M,\eta,i) & = & \frac{4\eta M}{d}\, \cos i\, v_F^{2}(t)\, \cos \phi(t),
\label{kt7}
\end{eqnarray}
with $v_F=(\pi M F)^{1/3}$ a velocity parameter, and the beam-pattern functions are
\begin{eqnarray}
F_+\left(\theta, \bar{\phi}, \bar{\psi}\right) &=& 
\frac{1}{2} \left( 1 + \cos^2 \theta\right) 
\cos 2\bar{\phi} \cos 2\bar{\psi} -  \cos \theta \sin 2\bar{\phi} \sin 2\bar{\psi},\\
\label{kt4}
F_\times\left(\theta, \bar{\phi}, \bar{\psi}\right) &=& 
\frac{1}{2} \left( 1 + \cos^2 \theta\right) 
\cos 2\bar{\phi} \sin 2\bar{\psi} +  \cos \theta \sin 2\bar{\phi} \cos 2\bar{\psi}.
\label{kt5}
\end{eqnarray}
Using the above expressions for the beam-pattern functions and
gravitational wave amplitudes the response takes the form
\begin{equation}
h(t) = \frac{4\eta M}{d} {\cal C}\, v_F^{2}(t)\, \cos[\phi(t) + \phi_0],
\end{equation}
where the amplitude coefficient ${\cal C}$ and phase $\phi_0$ can be assumed to 
be constant for signals lasting for a short duration (say, less than about 30 mins):
\begin{eqnarray}
{\cal C} & = & \frac{1}{2}\sqrt{\left( 1 + \cos^2 i\right)^2 F_+^2 + 4 (\cos i)^2 F_\times^2},\\
\phi_0 & = & \tan^{-1} \left [ \frac{2F_\times \cos i}{F_+ (1 + \cos^2 i)} \right ].
\label{kt8}
\end{eqnarray}
Thus, gravitational wave antennas are not able to extract the two polarisations
separately and the data analysis problem boils down to matching the time-varying
phase $\phi(t),$ and to a lesser extent the amplitude $v_F^{2},$ 
of the antenna response function.

Post-Newtonian theory and the quadrupole formula applied to a binary
give the relativistic binding energy 
$E(v_F)$ per unit mass, and the flux of the waves ${\cal F}(v_F)$ as series expansions 
in the parameter $v_F.$ 
Once we have the binding energy and the flux we can use the energy balance argument,
namely that the flux of gravitational waves is completely balanced by the
negative rate of change of the binding energy [$M\,dE(v_F)/dt = -{\cal F}(v_F)$], in order
to arrive at a parametrized equation for the evolution of the phase $\phi(t)$
of gravitational waves.  Integrating the energy balance equation supplemented 
by $2\pi F = d\phi/dt,$ one obtains:
\begin{eqnarray}
t(v_F) & = & t_{\rm ref} + M \int_{v_F}^{v_{\rm ref}} dv \, \frac{{E}'(v)}{{\cal F}(v)},
\label{eq:timeVsv} \\
\phi(v_F) & = & \phi_{\rm ref} + 2 \int_{v_F}^{v_{\rm ref}} dv\, v^3 \, \frac{{E}'(v)}{{\cal F}(v)},
\label {eq:phiVsv}
\end{eqnarray}
where $E'(v_F)=dE(v_F)/dv_F,$ $t_{\rm ref}$ is a reference time 
and $ \phi _{\rm ref}$ is  a reference phase of the signal, when the velocity parameter is $v_F=v_{\rm ref}.$
The numerical integration of the above equations is more economical when the following
differential equations are solved instead,
\begin{equation}
\frac{dv_F}{dt} = -\frac{{\cal F}(v_F)}{ME'(v_F)}, \ \ \ \ 
\frac{d\phi}{dt} = \frac{2 v_F^3}{M} \label{eq:TDODEs}.
\end{equation}
The parametric representation,  Eqs. (\ref{eq:timeVsv}) and (\ref{eq:phiVsv}), of the phasing
formula $\phi = \phi(t)$ holds under the assumption of `adiabatic inspiral',
{\it i.e.}, that gravitational radiation damping can be treated as
 an adiabatic perturbation of a circular motion. However, the effective
one-body approach  \cite {BD,Damour01,DIS3,DIJS1} has allowed a treatment of the
radiation damping to proceed beyond the adiabatic approximation.
In order to extract an inspiral signal that may be buried in noisy data
by the method of matched filtering, we need to employ  post-Newtonian accurate 
representations  for the two functions
${E}' (v_F)$ and ${\cal F}(v_F)$ that appear in the above phasing formulas.
Given an approximant $E_A (v_F)$, ${\cal F}_A (v_F)$,
one can define by replacing $E(v_F)\rightarrow E_A(v_F)$, 
${\cal F}(v_F)\rightarrow {\cal F}_A(v_F)$ in Eqs. (\ref{eq:timeVsv}) and (\ref{eq:phiVsv})
some approximate parametric representation, $t=t_A(v_F)$,
$\phi=\phi_A(v_F)$, and therefore a corresponding approximate template
\begin{equation}
h^A = h^A (t;\, {\cal C},\, t_{\rm ref},\, \phi_{\rm ref},\, M,\eta) \, , 
\label{4.3}
\end{equation}
obtained by replacing $v_F$, in the following $v_F$-parametric representation
of the waveform 
\begin{equation}
h^A (v_F) = \frac{4\eta M}{d} {\cal C} \, v_F^2 \, \cos \, \phi_A (v_F) \, , 
\label{4.4}
\end{equation}
by the function of time $v_F=v_A(t)$ obtained by inverting
$t=t_A(v_F)$. 

There are several ways of performing this inversion
which leads to the different T-approximants, 
{\it TaylorT1} [keep the rational polynomial $E'(v_F)/{\cal F}(v_F)$ in Eqs.~(\ref{eq:timeVsv}) 
and (\ref{eq:phiVsv}) as is and solve the integrals numerically], 
{\it TaylorT2} 
[re-expand the rational polynomial as a post-Newtonian series, truncate terms
to the appropriate order and solve the integrals analytically
to get series expansions, $t=\sum_k t_kv_F^k$ and $\phi=\sum_k \phi_kv_F^k,$
but solve numerically for $\phi(t)$ from the parametric equations] and 
{\it TaylorT3} [do as in TaylorT2 but also invert the series expansions
to obtain the phasing as an explicit function of time: $\phi(t) =
\sum_k \phi_k (t_{\rm ref}-t)^{-k/8}$], as discussed in 
Ref.~\cite{DIS3}.

It is often convenient to deal with the Fourier representation of
the waveform.  In the stationary phase approximation the 
Fourier transform, defined as $\tilde h(f) = \int_{-\infty}^\infty h(t) \exp(2\pi i f t) dt$,  
for positive frequencies reads~\cite{Thorne,SathDhur,DWS,DIS2}
\begin{equation}
\tilde{h}(f) \equiv \int_{-\infty}^\infty h(t) \exp(2\pi i f t)\, dt
=\frac{2\eta M {\cal C}}{d} \frac{v_f^2}{\sqrt{\dot{F}}}
e^{i\left[\psi(f)-\frac{\pi}{4}\right]},
\label{d4.6a}
\end{equation}
and, since $h(t)$ is real, $\tilde h(-f) = \tilde h^*(f).$
Here $t_f$ is the stationary point of the phase $\psi$ (i.e., $t_f$ is defined
by $d\psi/dt|_{t_f}=0$),
$v_f = (\pi M f)^{1/3},$ $\dot F$ is the time-derivative of 
the instantaneous gravitational wave frequency evaluated at the stationary
point given by, 
\begin{equation}
\dot F = -\frac{3v_f^2}{\pi M^2} \frac{{\cal F}(v_f)}{E'(v_f)},
\end{equation}
and $\psi(f)= 2\pi f t_f - \phi(t_f)$ is the phase of the 
Fourier transform in the stationary phase approximation given by
\begin{equation}
\psi(f) = 2\pi i f t_{\rm ref} - \phi_{\rm ref} + 2 \int_{v_f}^{v_{\rm ref}}
\left ( v_f^3 - v^3 \right ) \frac {E'(v)}{{\cal F}(v)}\, dv.
\end{equation}
Instead of solving the integrals in the above equation it is numerically
efficient to solve the following equivalent differential equations for
the phasing formula in the Fourier domain:
\begin{equation}
\frac{d\psi}{df} - 2\pi t = 0, \ \ \ \
\frac{dt}{df} + \frac{\pi M^2}{3v_f^2} \frac{E'(f)}{{\cal F}(f)} = 0.
\label {eq:frequency-domain ode}
\end{equation}
And in this study we have used the above differential equations in computing the
waveforms. For the energy we have used the exact analytical
formula (see next Section);  for the flux we have used the exact 
numerical flux to define the exact waveform and the perturbative 
expansions, or their re-summed improved versions, to define the
approximate waveforms.

\subsection{Signal-to-Noise Ratio} \label{sec:snr}
In order to estimate the signal-to-noise ratio (SNR) we shall assume that the signal
is detected using matched filtering and that the bank of templates used in
matched filtering has a waveform that matches the unknown signal very closely. 
To compute the optimum SNR we need to known the noise spectral density of the
detectors.  For initial LIGO, the one-sided noise power spectral density (PSD)
from the design study \cite{LIGO} is given by
\cite{DIS2}
\begin{equation}
S_{h}(f) = 9\times10^{-46}\left[ 0.52 + 0.16x^{-4.52} + 0.32 x^{2} \right]\ {\rm Hz}^{-1},
\end{equation}
where $x\equiv f/f_k,$ and $f_{k} = 150$\,Hz is the ``knee-frequency'' of the detector.  
We take the PSD to be infinite below the lower frequency cutoff of $f_{\rm low} = 40$\,Hz.  
For the VIRGO detector the PSD is given by
\begin{equation}
S_{h}(f) = 3.24\times10^{-46}\left[10^{3} (25 x)^{-5} + 2/x + 1 + x^{2} \right]\ {\rm Hz}^{-1},
\end{equation}
where $f_{k} = 500$\,Hz and $f_{\rm low} = 20$\,Hz.  Finally, for the GEO detector the
expected noise PSD is \footnote{Note that the expected PSD for GEO600 has changed over the
years. However, we continue to use the ``design'' sensitivity curve in order to make
it easier for comparing the results of this study with previous studies.}
\begin{equation}
S_h(f) = 10^{-46} \times \left [
(3.4 x)^{-30} + 34/x + 20(1-x^2+ x^4/2)/(1+x^2/2)
\right ]\  {\rm Hz}^{-1}
\end{equation}
where $f_k= 150$~Hz and the noise is assumed to be infinite below a lower frequency
cutoff of $f_{\rm low} = 40$~Hz.

Next, let us define the scalar product of two waveforms $h$ and $g$ by
\begin{equation}
\left<a\left|b\right.\right> 
=2\int_{0}^{\infty}\frac{df}{S_{h}(f)}\,\left[ \tilde{h}(f)\tilde{g}^{*}(f) +  \tilde{h}^{*}(f)\tilde{g}(f) \right],
\label{eq:scalarprod}
\end{equation}
where the * denotes complex conjugate and 
$\tilde{h}(f),\, \tilde g(f)$ are the Fourier transforms of $h(t),\, g(t)$.
The (square) of the signal-to-noise ratio obtained by matched filtering a 
signal $h$ with a template $g$ is given by
\begin{equation}
\rho^{2} \equiv \left(\frac{S}{N}\right)^{2} = 
\frac{\left<h\left|g\right.\right>^{2}}{\left<g\left|g\right.\right>} .
\end{equation}
If the template perfectly matches the signal then the SNR is simply 
$\rho^2 = \left <h,\, h\right >.$ For the inspiral signal given in Eq.~(\ref{d4.6a}),
and using the {\it Newtonian values} for the energy and flux functions, namely
$E' = -\eta v_f,$ ${\cal F} = 32\eta^2 v_f^{10}/5,$ and $\dot F = 96\eta v_f^{11}/(5\pi M^2),$
the SNR reduces to:
\begin{equation}
\rho(M, \eta, d; \theta, \bar\phi, \bar \psi, i) 
= \frac {M^{5/6}\eta^{1/2}}{d\pi^{2/3}} \,
{\cal C}(\theta, \bar\phi, \bar\psi, i)
\left [ \frac{5}{6} \int_0^{F_{\rm cut}}
\frac{f^{-7/3}}{S_h(f)}\, df \right]^{1/2},\ \
\end{equation}
Here $F_{\rm cut}$ is the frequency where the data analyst chooses to terminate
the signal. The cutoff frequency is usually chosen to be the frequency of the
waves at the last stable orbit of the binary (cf.~Sec.~\ref{subsec:energy}) which depends 
on the total mass of the two stars and their spin magnitude.
Note that the SNRs depend only on the combination ${\cal M}=M\eta^{3/5}$ -- the chirp mass
-- and not on individual masses of the system.
For a given total mass the SNR at a fixed distance (alternatively the span of the antenna
at a fixed SNR) is the highest for binaries of equal masses (when
$\eta=1/4$) and reduces by the fraction $\sqrt{4\eta}$ for binaries of unequal masses.
The SNR decreases inversely as the distance to the source and
depends not only on the intrinsic parameters $M,$ $\eta$ and $F_{\rm cut},$ 
but also on the relative orientations of the source and the antenna specified
by the angles $\theta,$ $\bar\phi,$ $i,$ $\bar \psi.$ We can compute the SNR for 
a source of RMS orientation by averaging the amplitude factor ${\cal C}$ over all 
the angles.  Performing the average over the angles leads to: 
\begin{equation}
\langle F_+^2\rangle_{\theta,\bar{\phi},\bar{\psi}} =  
\langle F_\times^2\rangle_{\theta,\bar{\phi},\bar{\psi}} =  \frac{1}{5}, \ \ \ \ 
\langle {\cal C}^2\rangle_{i,\theta,\bar{\phi},\bar{\psi}} =  \frac{4}{25}.
\label{kt6}
\end{equation}
For an ideally oriented source, that is for a source that produces the highest
SNR, ${\cal C} = 1.$ Thus, the RMS and ideal SNRs,
obtained by using ${\cal C}=2/5$ and ${\cal C}=1,$ respectively, are given by
\footnote{The expression for the SNR in this paper differs from that in
Ref.~\cite{DIS2} by a factor $\sqrt{2}$ because here we use one-sided PSD
as opposed to two-sided PSD used in Ref.~\cite{DIS2}.}
\begin{equation}
\rho_{\rm RMS} = \frac {{\cal M}^{5/6}}{d\pi^{2/3}} 
\left [\frac{2}{15} \int_0^{F_{\rm cut}} 
\frac{f^{-7/3}}{S_h(f)}\, df\right]^{1/2},\ \
\rho_{\rm ideal} = \frac{5}{2} \rho_{\rm RMS}.
\label{final snr}
\end{equation}

\begin{figure}[t]
\begin{center}
\includegraphics[width=2.2in]{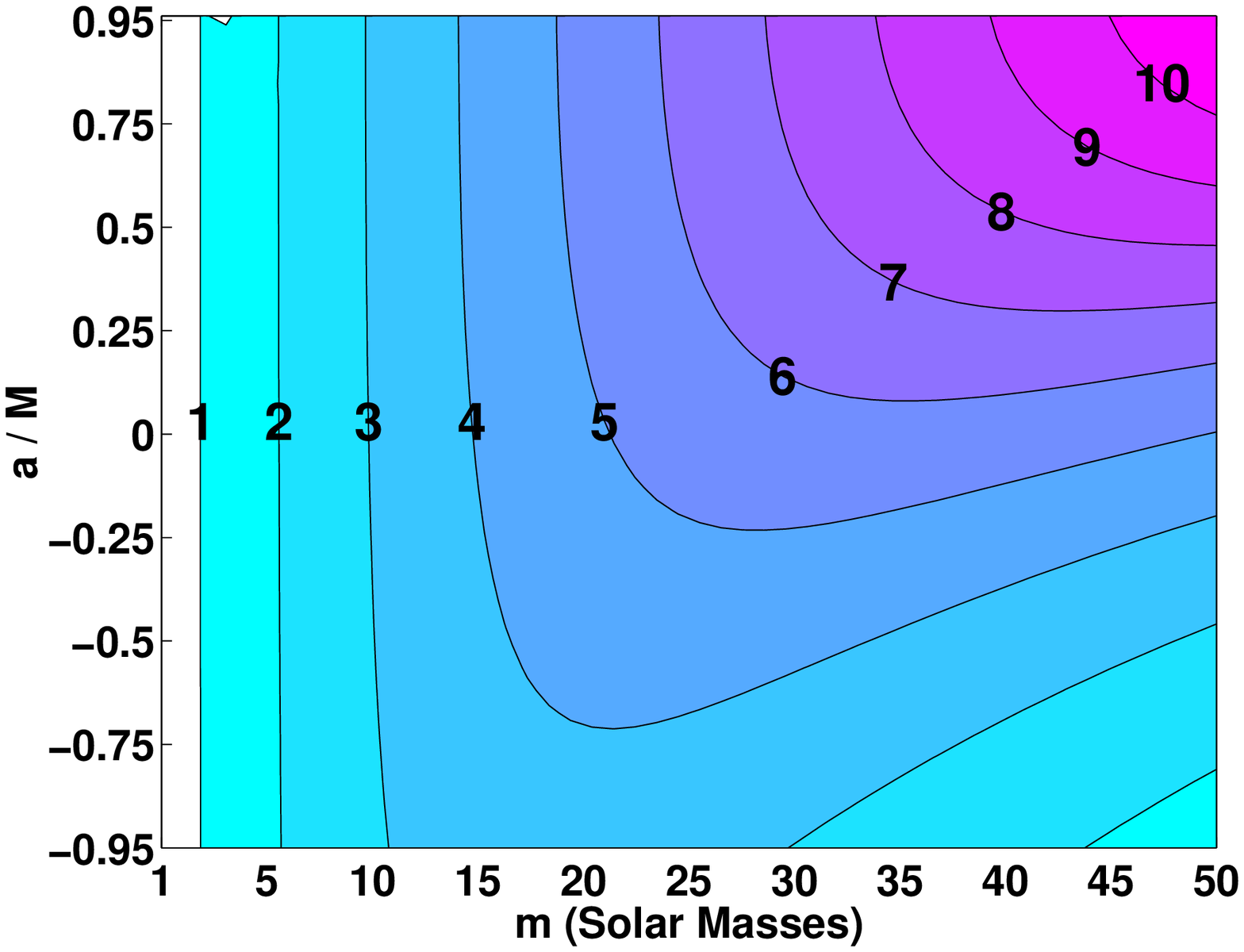}
\includegraphics[width=2.2in]{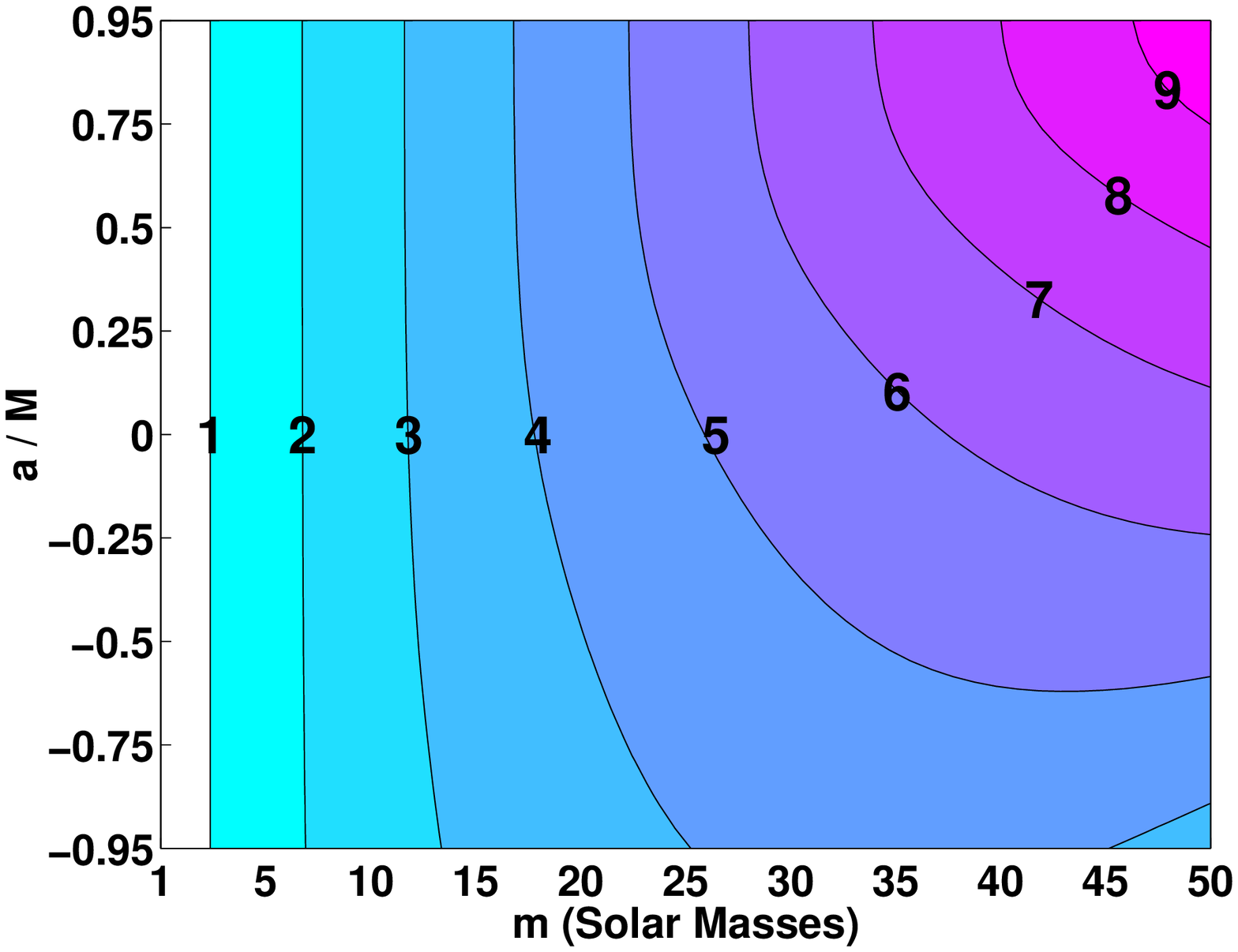}
\includegraphics[width=2.2in]{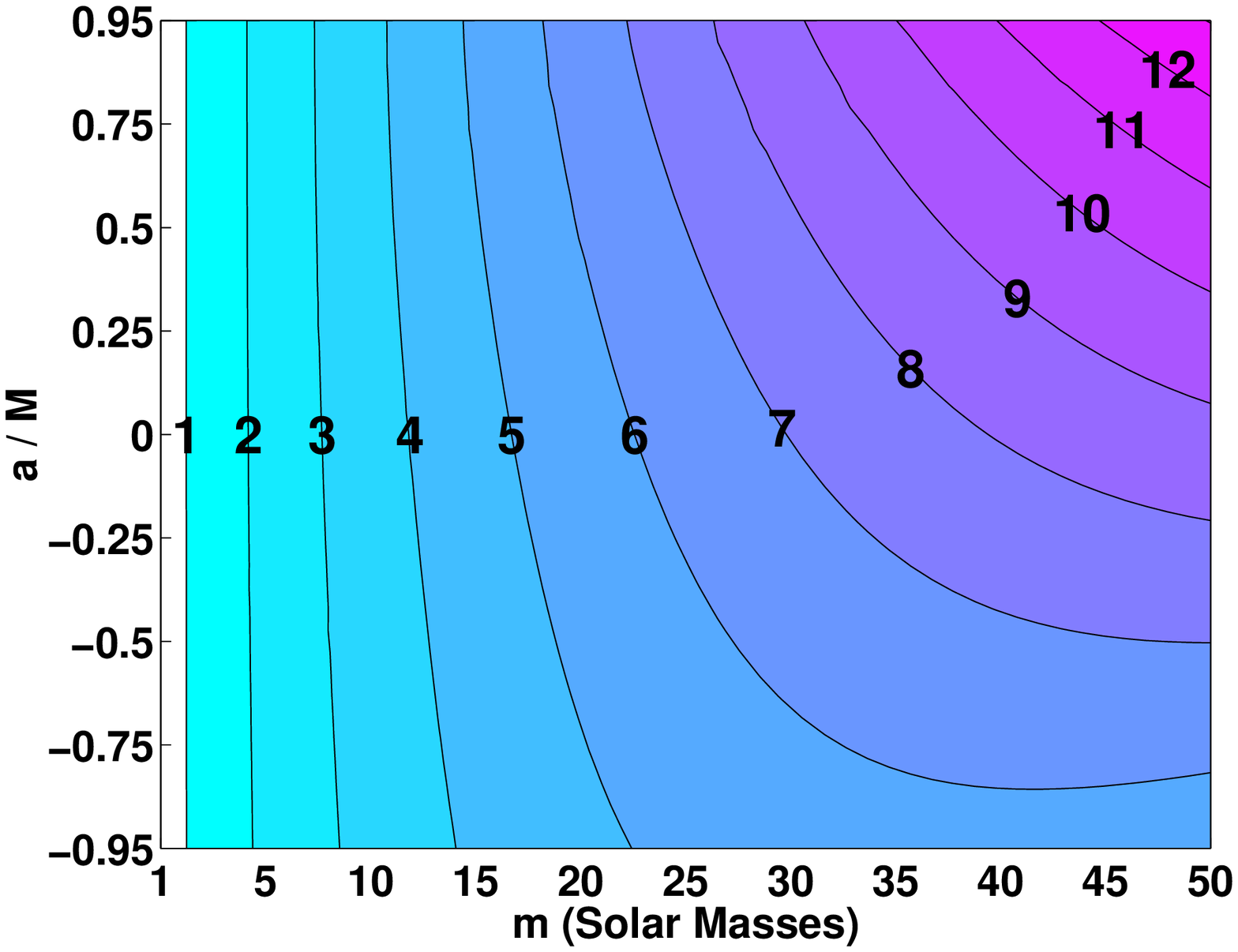}
\caption{The RMS SNR for equal mass binaries at 100 Mpc for the LIGO (left), GEO (center)
and VIRGO (right) antennas.}
\label{fig:snr}
\end{center}
\end{figure}

In Fig.~\ref{fig:snr} we have plotted the {\it RMS} SNR 
for the LIGO, VIRGO and GEO detectors, as a function of the spin magnitude
and the total mass of the system.  
Clearly, systems with large spins in which the test mass is in prograde orbit
produce the highest SNRs since their last stable orbit frequency is higher
than those of retrograde orbits and they last longer. To capture such signals, 
however, one must know the phasing of the waves very well
since the bodies spend a substantial fraction of their time in the detector
band in the strongly non-linear regime of the evolution. We shall show later in this paper that
there could be a significant drop in the SNR, and a corresponding drop in
the number of candidate events that can be detected, while P-approximants
lose little SNR and therefore capture almost all events. 

\section{Gravitational Binding Energy and Flux Functions}
\label{sec:EnergyAndFlux}
We can see from Eqs.~(\ref{eq:timeVsv}) and (\ref{eq:phiVsv})
that the phase of the gravitational wave depends both on 
the energy and flux functions of the binary system.  In 
the test-mass case, that is when one of the masses is 
far smaller than the other ($4\eta \ll 1$) so that the dynamics is
governed entirely by the Kerr geometry of the massive body, 
we have an exact expression for the energy $E(x)$, but only a series 
representation for the flux $F(x)$.  In the comparable-mass case, wherein one cannot
neglect the perturbation caused by the companion, both functions are 
represented only by post-Newtonian expansion.  The standard approximants for 
$E(v)$ and ${\cal F}(v)$ are simply  their successive Taylor approximants  
$E_{T_n}$ and ${\cal F}_{T_n},$ respectively. The gravitational wave signal
constructed using these Taylor approximants are called T-approximant signals
or waveforms are simply T-approximants.

It was shown in Ref~\cite{DIS1} that the convergence properties of 
both the energy and flux functions can be improved by using 
Pad\'e approximation of well-defined energy and flux functions.
Damour, Iyer and Sathyaprakash followed a two-pronged approach to
construct improved energy and flux functions \cite{DIS1}: 
Starting from the more basic energy-type and flux-type functions,  $ e(v)$
and $l(v),$ Ref.~\cite{DIS1} constructs Pad\'e-type approximants, 
say $e_{P_n}$, $l_{P_n}$, of the ``basic'' functions $ e(v)$, $l(v).$
One can then compute the required energy and flux functions entering the 
phasing formula.  The successive approximants
$ E[e_{P_n}]$ and  $ {\cal F}[e_{P_n} , l_{P_n}]$ 
have better convergence properties than  their Taylor counterparts 
$ E_{T_{n}}[e_{T_n}]$ and  $ {\cal F}_{T_{n}}[e_{T_n} , l_{T_n}]$. 

In this study we will restrict ourselves to the test mass approximation and, therefore,
employ the exact energy function. However, in the same approximation
the flux function has been computed analytically only as a Taylor series 
although numerically the flux has been computed exactly. Since our aim
is to draw conclusions on how effective P-approximants are in the comparable
mass case, wherein one has only a Taylor expansion of the flux, we construct
P-approximants of flux and compare it with the numerical results.
The reasons for constructing a new flux function are the following: 
Firstly, it is well known, in the Schwarzschild case, that a simple pole 
exists at $r=3\,M$~\cite{Poisson1} in the expression for the flux.  
We might, therefore, expect that a similar pole exists in the Kerr case.  
A Taylor approximant of the flux function will never produce a pole 
while the equivalent Pad\'e approximant, which is essentially a rational
polynomial, will have
a pole.  Although it is well known that Pad\'e approximants have the 
ability to model functions which are subject to 
singularities~\cite{NumRec,BenOrz} it is not guaranteed that   
the rational polynomial will reproduce the exact pole.  Secondly, it is often possible 
to recover from the {\it divergence}~\footnote{The divergence referred to here is the 
behaviour of the successive orders of the Taylor expansion of the flux
which might not converge as we go to higher orders.}
of a Taylor series by using Pad\'e approximation.  
Again, one cannot be certain that the resulting Pad\'e approximant, even when
it converges,  will be closer to being exact than any of the Taylor approximants, but
experience with the test mass approximation in the Schwarzschild case 
render some optimism to this expectation.

We create a Pad\'e approximation of a truncated power series of an analytic
function containing $n$ terms by equating it to a rational function 
such that the rational function when expanded as a power series and truncated
to order $n$ coincides with the original power series. More precisely,
Pad\'e approximation can be thought of as an operator $P_M^N$ 
that acts on a polynomial $\sum_{k=0}^n a_k v^k$ to define a 
rational function:
\begin{equation}
\newcommand\Dfrac[2]{\frac{\displaystyle #1}{\displaystyle #2}}
P^{N}_{M} \left (
\sum_{k=0}^{n} a_{k}v^{k} \right )
= \Dfrac{\sum_{k=0}^{N} A_{k}\,v^{k}}{\\ 
1+\sum_{k=1}^{M} B_{k}\,v^{k}},
\label{eq:pade}
\end{equation}
such that the number $ N + M + 1$ of coefficients  in the rational polynomial
on the right hand side is the same as the number $n+1$ 
of Taylor coefficients  on the left hand side.
By setting $N=M + \epsilon$ with 
$\epsilon = 0,\ 1,$ we can define two types of Pad\'e approximants: 
These are the super-diagonal, $P^{M+\epsilon}_{M}$, and sub-diagonal, 
$P^{M}_{M+\epsilon}$, approximants.  Normally, the 
sub-diagonal approximants are preferred over super-diagonal approximants.  
This is because when $M=N+\epsilon$ 
the right hand side of Eq.~(\ref{eq:pade}) can be re-expanded as a 
continued fraction which have the property that as 
we go to each new order of the power series only one new coefficient 
needs to be calculated.  Conversely, with the super-diagonal approximants, 
we would have to re-calculate all the $A$'s and $B$'s in the above 
equation as we go to higher orders in the Taylor expansion.  This
means that the sub-diagonal Pad\'e approximants are more {\it stable} and if we
see a trend of convergence in the coefficients the addition of a term
is not likely to spoil this convergence.
We refer the reader to Appendix~\ref{sec:padeapp} for a more 
summary of the properties of Pad\'e approximations and 
how to find the diagonal Pad\'e coefficients using continued fraction 
expansion.


\subsection{The Orbital Energy.}\label{subsec:energy}
In the case of both Schwarzschild and Kerr black holes we have an exact 
expression for the orbital energy of a test particle in a circular orbit around 
the parent black hole.  For a black hole of mass $M$ the energy $E$
in terms of the dimensionless magnitude of velocity $v 
\equiv \sqrt{M/r},$ $r$ being is the radial coordinate in the Boyer-Lindquist 
coordinates, takes the form~\cite{Bardeen}
\begin{equation}
E(v, q) = \eta\,\frac{1-2\,v^{2}+q\,v^{3}}{\sqrt{1-3\,v^{2}+2\,q\,v^{3}}}.
\label{eq:energy}
\end{equation}
where $q$ is a dimensionless spin parameter given in terms of the spin
angular momentum $J$ of the black hole by $q \equiv J/M^2 \equiv a/M,$ 
with $a$ spin angular momentum per unit mass in the Kerr metric.
It is actually the derivative of the orbital 
energy that appears in the phasing formula given by:
\begin{equation}
E'(v, q) = -v\,\eta\,\frac{1-6\,v^{2}+8\,q\,v^{3} - 
3\,q^{2}\,v^{4}}{\left(1-3\,v^{2}+2\,q\,v^{3}\right)^{3/2}},
\label{eq:energyderiv}
\end{equation}
where a prime denotes a derivative with respect to the velocity parameter $v$.  

For black holes with spin, the positions of the last stable circular orbit is a 
function of the spin parameter $q$.  In order to find the position of the last 
stable circular orbit we take $E'(v, q)=0$ which gives~\cite{Bardeen}:
\begin{equation}
r_{LSO}^{\pm}(q) =M\,\left[ 3 + z_{2}(q)\mp \sqrt{\left[ 3 - 
z_{1}(q)\right]\,\left[ 3 + z_{1}(q) + 2\,z_{2}(q) \right]}\right],
\end{equation}
where
\begin{eqnarray}
z_{1}(q) & = & 1 + \left(1 - q^2\right)^{\frac{1}{3}}\left[\left(1 + q\right)^{\frac{1}{3}} + \left(1 - 
q \right)^{\frac{1}{3}}\right]\\ \nonumber \\
z_{2}(q) & = & \sqrt{3\,q^2 + z_{1}^{2}}.
\end{eqnarray}
The $+$ $(-)$ sign on $r_{LSO}$ corresponds to prograde (retrograde) orbits.
Now the position of the last unstable circular orbit or ``photon 
ring'' occurs where the energy function exhibits a pole. Thus, 
the photon orbit is found by solving a cubic equation.  Of the 
three poles, the physically relevant pole is given by~\cite{Bardeen,Chandra}
\begin{equation}
r_{pr}^{\pm}(q) = 2\,M \left[ 1 + \cos\left[ \frac{2}{3}\,\cos^{-1}\left(\mp q \right)\right]\right],
\end{equation} 
where once again the different signs define a prograde (retrograde) orbit.  In 
the limit of $q\rightarrow 1$, both $r_{LSO}$ and $r_{pr}$ move towards the 
horizon of the black hole, 
\begin{equation}
r^{\pm}_{H} = M\left (1 \pm \sqrt{1 - q^{2}}\right ).
\end{equation}
In the maximally rotating case of $q=1$, we cannot distinguish between the last
stable orbit and the light ring.
This means, the greater the spin of the BH, the closer the 
particle can approach to it before beginning the plunge phase.  On the other 
hand, in the limit $q\rightarrow -1$, the position of $r_{LSO}$ moves outwards 
from the BH.  Thus the particle begins its plunge much earlier than in the 
prograde case.  Finally, for a particle in an
orbit about a Schwarzschild black hole the above equations give the 
familiar position of $r=6\,M$ for the LSO and 
$r = 3\,M$ for the photon ring.


\subsection{The Flux Function}\label{subsec:flux}
As discussed in the Introduction it has not been possible to derive
an exact analytical expression for the flux of gravitational waves 
emitted by a binary system although analytic approximate expressions, and exact
numerical results, have been computed.
In the interesting case of two comparable masses in orbit around each
other post-Newtonian methods have been used to derive an expression for
the flux to quite a high order in the expansion parameter. However, since post-Newtonian
theory is known to be poorly convergent, especially when the expansion parameter
approaches unity, it has been suggested to employ in its place an equivalent
rational polynomial, or Pad\'e approximation, to the (modified) flux function. 
Since the purpose of this paper is to test the effectiveness of such an 
approximation we need a firm ground on which we can conduct our test.

In the case of a test mass in an equatorial orbit around a Kerr black hole 
numerical methods have been used to compute the flux to all post-Newtonian orders.
This is, of course, valid only in the limit of vanishingly small mass of the
test body as compared to the central object. However, all the relativistic corrections,
including hereditary effects, such as the back scattering of gravitational waves off
the curved background geometry, would be present in this computation. 
The deformation of the geometry due to the presence of the
second body, or the back reaction of the waves on the motion of the test body,
which would in turn affect the emission process, will not have been included
in such a calculation.  The deformation of the geometry, parametrized
by the symmetric mass ratio of the system $\eta,$
is important only a few orbits before the two objects merge. This should be
expected from the fact that both in Newtonian and Einsteinian gravity it has
been possible to construct an effective one-body formalism to describe the
dynamics of a binary and the dynamics derived within this formalism is expected
to be valid close to the point when the two bodies plunge towards each other. 
Therefore, we expect that the deformation of the system will not bring
about a major change in the analytic behaviour of the flux and lessons
learnt in the test mass approximation will be applicable, albeit qualitatively,
in the comparable mass case.

Thus, our strategy is to employ the test mass exact flux computed numerically
together with the analytic expression for the energy function discussed 
in the previous Section. These two exact functions, together with the energy
balance equation, can be used to construct an exact phasing formula for the
inspiral signal. We shall also define an approximate phasing formula using
the corresponding post-Newtonian expansions of the two functions. Our
approximations at each post-Newtonian order will be one of two types: (1) The
standard post-Newtonian or (2) the improved Pad\'e approximation. In the next
Section we will discuss how these approximations can be further improved and
in the last Section we will measure the quantitative performance of the two
approximations.

\subsection{Post-Newtonian flux function}

For a test-particle in a circular equatorial orbit the
post-Newtonian expansion of the flux function has been calculated 
up to ${\mathcal O}$ ($v^{11}$) in the case of a Schwarzschild BH \cite{TTS},
and to ${\mathcal O}$($v^8$) in the case of a Kerr BH \cite{SSTT,TSTS}.  
The general form of the flux function in both these cases is given by the 
expression
\begin{equation}
F_{T_{n}}(x;\, q) = F_{N}(x)\left[\sum_{k=0}^{n}\,a_k(q)x^{k} + 
\ln(x)\,\sum_{k=6}^{n}\,b_{k}(q) x^{k} + {\mathcal O}\left(x^{n+1}\right)\right].
\label{eq:flux}
\end{equation}
where the expansion is known to order
$n = 8$ and 11, in the Kerr and Schwarzschild cases, respectively, 
$q$ is the dimensionless spin parameter introduced earlier 
and $F_{N}(x)$ is the dominant {\it Newtonian} flux function given by
\begin{equation}
F_{N}(x) = \frac{32}{5}\eta^{2}x^{10}.
\end{equation}
Here, $x$ is the magnitude of the invariant velocity parameter observed 
at infinity which is related to the angular frequency $\Omega$ by $x = 
\left(M\Omega\right)^{1/3}$.  The relation between the parameter $x$ and the 
local linear speed $v$ in Boyer-Lindquist coordinates is given by
\begin{equation}
x(v, q) = v\left[1-qv^{3} + q^{2}v^{6} \right]^{1/3},
\end{equation}
which reduces, in the Schwarzschild limit, to $x=v.$ Note that this local linear velocity 
is related to the Boyer-Lindquist radial coordinate $r$ by $v^2 = M/r.$ The spin-dependent 
coefficients, $a_{k}(q)$, and the log-term coefficients, $b_{k}$, are 
given by\footnote{Note that the post-Newtonian coefficients of flux in 
are taken from Ref.~\cite{TSTS} Eq.~(G19).
We have confirmed with Tagoshi et al. that the
term $359\pi q/14$ has the wrong sign in some later references but Ref.~\cite{TSTS}
has the correct sign.},
\begin{eqnarray}
& a_{0} = 1, \,\,\,\,\,a_{1} = 0,\,\,\,\,a_{2}=-\frac{1247}{336}, \,\,\,\,a_{3} = 
4\pi-\frac{11\,q}{4}, 
\,\,\,\,a_{4}=-\frac{44711}{9072}+\frac{33\,q^{2}}{16},\,\,\,\,a_{5} = 
-\frac{8191\,\pi}{672}-\frac{59\,q}{16} &
\nonumber \\ 
& a_{6} = 
\frac{6643739519}{69854400}-\frac{1712\,\gamma}{105}+\frac{16\,\pi^{2}}{3}-\frac
{3424\,\ln 2}{105}-\frac{65\,\pi q}{6} + 
\frac{611\,q^{2}}{504},&
\nonumber \\ 
& a_{7} = -\frac{16285\,\pi}{504} + \frac{162035\,q}{3888} + 
\frac{65\,\pi\,q^{2}}{8} - 
\frac{71\,q^{3}}{24},&
\nonumber \\ 
& a_{8} = -\frac{323105549467}{3178375200} + \frac{232597\,\gamma}{4410} 
-\frac{1369\,\pi^{2}}{126} +\frac{39931\,\ln 2}{294} -\frac{47385\,\ln 3}{1568} 
-\frac{359\,\pi\,q}{14} +\frac{22667\,q^{2}}{4536} +\frac{17\,q^{4}}{16},& 
\nonumber \\
& b_{6} = -\frac{1712}{105},\,\,\,\,b_{7} = 0, \,\,\,\,b_{8} = 
\frac{232597}{4410}, &
\end{eqnarray}
where $\gamma$ is Euler's number.  For graphical purposes, it is more 
appropriate to deal with the Newton-normalized fluxes
defined by 
\begin{equation}
\widehat{F}_{T_{n}}(x) \equiv F_{T_{n}}(x) / F_{N}(x).
\end{equation}

\subsection{P-approximant of the flux function}\label{sec:p-approximant}
We will now outline the method of calculation the P-approximant of the 
GW flux as proposed by DIS~\cite{DIS1}.  We notice from the form of the series 
expansion for the flux, Eq.~(\ref{eq:flux}), that we begin to encounter 
logarithmic terms at order $x^{6}$ and above.  In general, series 
approximations of this form have slow convergence properties.    In order to 
prepare the series representation of the flux for creating the Pad\'e 
approximation, it is convenient if we factor out the logarithmic terms.  We can 
then write Eq.~(\ref{eq:flux}) as
\begin{equation}
\widehat{F}_{T_{n}}(x) = 
\left[1+\ln\left(\frac{x}{x_{LSO}}\right)\sum_{k=6}^{n}\,l_{_{k}}x^{k}\right]
\left[\sum_{k=0}^{n}\,c_{_{k}}x^{k}\right],
\end{equation}
where the new coefficients $c_{k}$ and $l_{k}$ are functions of the old 
coefficients $a_{k}$ and $b_{k}.$ As in Ref.~\cite{DIS1} the log-terms 
have been ``normalized'' using the value of the velocity parameter 
at the LSO; this helps in reducing the importance of the log-terms.  
Factoring out the logarithmic terms aids
in constructing the rational polynomial, or the Pad\'e approximation, of the
reminder. Moreover, since we expect the flux to have a pole at the location
of the light ring it is best to factor out the expected pole so that the
reminder  has good analytical properties. To this end
we create the factored flux function, $f_{T_{n}}(x)$ by the operation
\begin{equation}
f_{T_{n}} \equiv \left(1-\frac{x}{x_{pole}}\right)\widehat{F}_{T_{n}}.  
\end{equation}
Factoring out the pole also helps to alleviate the problem arising from
the absence of the linear term in the PN expansion of the flux. (Note that
$a_1=0$ in both the Schwarzschild and Kerr cases.)
We can see from Appendix~\ref{sec:padeapp} that in the absence of such a
term the continued fraction form of the  Pad\'e approximation, the so-called
diagonal Pad\'e approximant, would lead either to zero or infinite 
Pad\'e coefficients.  The above operation rectifies this problem by introducing 
a linear term into the Taylor series for the flux. If we write the expression 
in full we obtain
\begin{equation}
f_{T_{n}}(x) = 
\left[1+\ln\left(\frac{x}{x_{LSO}}\right)\sum_{k=6}^{n}\,l_{_{k}}x^{k}\right]
\left[\sum_{k=0}^{n}\,f_{_{k}}x^{k}\right],
\label{eq:fluxf}
\end{equation}
where $f_{_{0}} = c_{_{0}}$ and $f_{_{k}} = c_{_{k}} - c_{_{k-1}}/x_{pole},$ 
$k = 1,\ldots,n$.  

We can now construct a new flux function by using the Pad\'e
approximant of the factored flux given above.  Indeed,
we can construct two variants of the new flux: 
The first one is what we call the {\it Direct} or D-Pad\'e approximant, 
which is obtained by 
directly starting from the flux function $f(v)$ in Eq.~(\ref{eq:fluxf}) 
and constructing the equivalent rational polynomial. This is the approach followed
in DIS. An alternative approach to this is motivated by the fact that in the gravitational
wave phasing formula the flux appears in the denominator. Thus, instead of constructing
the Pad\'e approximant of the flux function one could first construct the polynomial expansion
of the inverse of the flux function and construct the Pad\'e approximant of the resulting
polynomial.  We call the approximant constructed this way as {\it  Inverse-} or I-Pad\'e 
approximant because it is obtained from the Taylor expansion of the {\it inverse} 
of the flux function $f(v)$ in Eq.~(\ref{eq:fluxf}).
Thus, our two improved versions of the flux are defined as follows:
The Direct Pad\'e approximant is defined by 
\begin{equation}
f_{DP_{n}}(v) \equiv 
\left[1+\ln\left(\frac{x}{x_{LSO}}\right)\sum_{k=6}^{n}\,l_{_{k}}x^{k}\right]P^{
m}_{m+\epsilon}\left[\sum_{k=0}^{n}\,f_{_{k}}x^{k}\right],
\end{equation}
where $P^{m}_{m+\epsilon}$ is the diagonal or sub-diagonal Pad\'{e} 
approximant, $n = 2m+\epsilon,$ with $\epsilon = 0$ or $1,$ depending on
whether $n$ is even or odd.  The Inverse Pad\'e approximant of flux is defined by
\begin{equation}
f_{IP_{n}}(v) \equiv 
\left[1-\ln\left(\frac{x}{x_{LSO}}\right)\sum_{k=6}^{n}\,l_{_{k}}x^{k}\right]P^{
m}_{m+\epsilon}\left[\sum_{k=0}^{n}\,d_{_{k}}x^{k}\right],
\end{equation}
where the coefficients $d_k$ in the Taylor expansion are defined by 
\begin{equation}
\sum_{k=0}^{n}\,d_{_{k}}x^{k} \equiv \left(\sum_{k=0}^{n}\,f_{_{k}}x^{k}\right)^{-1}.
\end{equation}
One finds $d_k$ by first expanding the RHS in a binomial series and then
identifying the coefficients of the various terms with those on the LHS.

Having constructed the rational polynomials equivalent to a given truncated
Taylor series of the modified flux function we can return to the original
flux function that appears in the phasing formula. We shall call the 
flux function so constructed as \emph{P-approximant,} and not just Pad\'e
approximant to remind ourselves that the new flux has been obtained by
improving the convergence properties in two steps (i.e., definition of a
new flux function and the construction of its rational polynomial) 
and not just a direct application of Pad\'e approximation.
Thus, we define the {\it Direct P-approximant} of flux as
\begin{equation}
\widehat{F}_{DP_{n}}(v) = 
\left(1-\frac{x}{x_{pole}}\right)^{-1}\,f_{DP_{n}}(v),
\end{equation}
and {\it Inverse P-approximant} as,
\begin{equation}
\widehat{F}_{IP_{n}}(v) = 
\left[\left(1-\frac{x}{x_{pole}}\right)\,f_{IP_{n}}(v)\right]^{-1}.
\end{equation}
Detailed investigation shows that 
Inverse P-approximants of flux, namely $\widehat{F}_{IP_n},$ have better convergence 
properties than Direct P-approximants. We shall therefore use
only $\widehat{F}_{IP_n}$ in all our investigations in the rest of this paper.


\subsection{Convergence of the Kerr flux function}\label{sec:fluxmodel}
In this Section we will look at the convergence properties of 
the post-Newtonian series representations and the improved 
P-approximants of the flux
function discussed in the last Section. To this end we shall 
employ all the post-Newtonian orders up to which the Taylor expansions have
been worked out in the Kerr case.  We shall investigate the 
convergence of the approximants in the case of a test particle 
(i.e., $\eta \rightarrow 0$ limit) either co-rotating 
(i.e., $q > 0,$ {\it prograde orbits}) or counter-rotating 
(i.e., $q < 0,$ {\it retrograde orbits}), with respect to a 
central black hole whose spin parameter $q$ takes nine values
over the full range from $(-1,\, 1).$ In this Section
we shall make a graphical comparison of the analytical and numerical
fluxes and assess the performance of the flux with respect to variation
in the spin-parameter values at increasingly higher post-Newtonian orders.
We will find that neither the post-Newtonian series nor their improved
versions fit the numerical flux exactly when the parameters are chosen
to be the same for analytical fluxes as the numerical flux. 
However, the P-approximant fluxes agree with the numerical 
fluxes quite closely when their spin values are slightly mis-matched
with the true values of the numerical fluxes. The post-Newtonian approximants
do show this improved behaviour, albeit not to the extent of
P-approximants.  Thus, we can expect the P-approximants to define
better signal models than post-Newtonian approximants with regard
to both faithfulness and effectualness.


\begin{figure}[t]
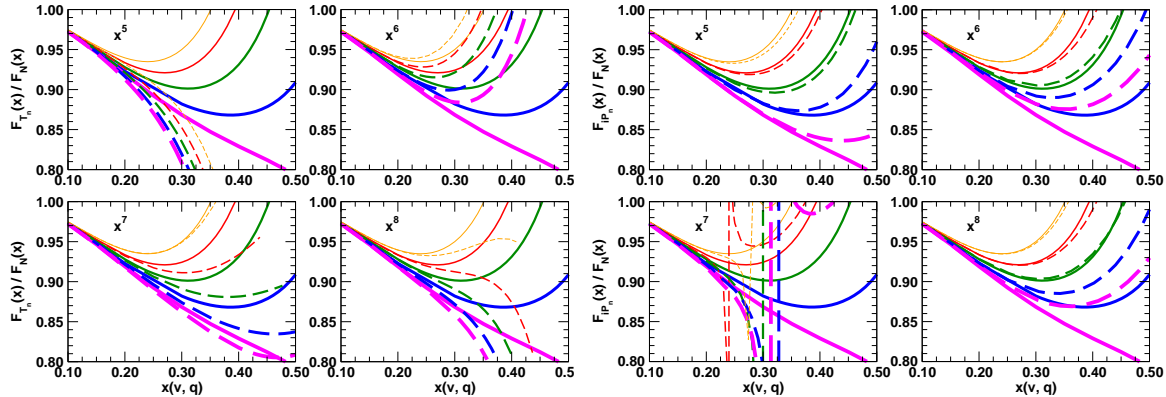

\begin{center}
\includegraphics[width=3in]{NNKpnfluxpap.eps}
\includegraphics[width=3in]{NNKipaprofluxpap.eps}
\caption{The convergence of the Newton-normalized T-approximants 
(left panels) and Inverse P-approximants (right panels) to
GW flux at post-Newtonian orders ${\cal O}(x^{5})$ to ${\cal O}(x^{8}),$ 
for a test particle in prograde orbit around a 
Kerr black hole. The solid lines represent the exact 
numerical fluxes with thicker lines corresponding to larger spin magnitudes,
$q = \{0,\, 0.25,\, 0.50,\, 0.75,\, 0.95\},$ respectively.
The correspondingly thickened dashed lines represent the Taylor- and P-approximants.}
\label{fig:KTP}
\end{center}
\end{figure}

\subsubsection{Prograde Orbits, $q \ge 0$}\label{subsec:promod}
The numerical fluxes have been computed exactly~\cite{Shibata}, but in the
test mass approximation, at a set of test values of
the spin parameter $q.$ For prograde orbits the flux has been computed at 
$q = \{0.25,\, 0.50,\, 0.75,\, 0.95\}$.  For the sake of
completeness, we shall also include in our comparison
the numerical flux for a particle in orbit about 
a Schwarzschild BH, i.e. $q=0$~\cite{TTS}.  In Fig.~\ref{fig:KTP}
we have plotted the Newton-Normalized numerical fluxes alongside 
analytical fluxes for both the truncated Taylor approximants 
(four panels on the left) and (Inverse) P-approximants 
(four panels on the right). For each approximant, we consider
four post-Newtonian orders at \{$x^5,\, x^6,\, x^7,\, x^8$\} and
five spin values.  The solid lines with increasing thickness
correspond to the exact numerical fluxes at progressively larger
values of the spin parameter starting at $q = 0,$ (thinnest line)
and ending at $q=0.95$ (thickest line). 
The corresponding dashed lines represent the analytical 
approximant to the GW flux.  

Let us first concentrate on the post-Newtonian fluxes:
We notice immediately that at all values of $q$, and large values of $x,$
the 2.5 post-Newtonian approximant ${\cal O}(x^{5})$ is highly divergent.
Indeed, this order has been known to exhibit improper behaviour in the
Schwarzschild case and continues to be so for prograde orbits in Kerr.
The situation improves at ${\cal O}(x^{6}),$ but begins to worsen again 
at ${\cal O}(x^{7}),$ becoming highly divergent at ${\cal O}(x^8)$.  The
plots serve to illustrate the principal problem with the PN expansion: 
Going to higher orders of the approximation does not necessarily 
correspond to a better accuracy.  

The sequence of panels on the right-hand-side of Fig.~\ref{fig:KTP} 
demonstrates that the P-approximant flux has improved
convergence properties over the post-Newtonian flux
for the same range of post-Newtonian orders and spins.  
The extreme divergence which the post-Newtonian flux 
exhibited at ${\cal O}(x^{5})$ and ${\cal O}(x^{8})$ 
has been cured.  Let us note, however, that sometimes the rational polynomial
approximation of a Taylor series can produce spurious poles 
in the region of interest~\footnote{Indeed, it has been known \cite{DIS1,DIJS1}
that occasionally spurious poles occur very close to the zeroes of 
the function so much so that the two could cancel each other but for a 
very slight difference in their locations. In such cases one cure would be
to remove both the pole and the zero by hand if they occur sufficiently
close to each other.}.
An example of this can be seen at ${\cal O}(x^7)$ which shows a spurious
pole at every value of $q.$
The main factor that stands out in the right panels of 
Fig.~\ref{fig:KTP} is the following: The P-approximant fluxes display 
the same characteristic shape as the numerical fluxes although they 
don't agree with the exact flux at higher values of the parameter $x.$  


\begin{figure}[t]
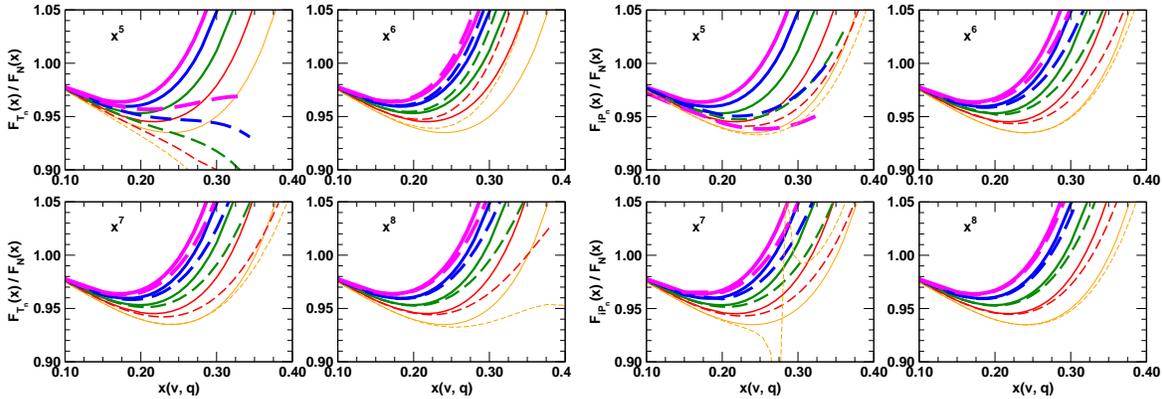

\begin{center}
\includegraphics[width=3in]{NNKpncrfluxpap.eps}
\includegraphics[width=3in]{NNKipacrfluxpap.eps}
\caption{The convergence of the Newton-normalized T-approximants 
(left panels) and Inverse P-approximants (right panels) to
GW flux at post-Newtonian orders ${\cal O}(x^{5})$ to ${\cal O}(x^{8}),$ 
for a test particle in retrograde orbit around a 
Kerr black hole. The solid lines represent the exact 
numerical fluxes with thicker lines corresponding to larger spin magnitudes,
$q = \{0,\, -0.25,\, -0.50,\, -0.75,\, -0.95\},$ respectively.
The correspondingly thickened dashed lines represent the Taylor- and P-approximants.}
\label{fig:KTR}
\end{center}
\end{figure}
In the case of prograde orbits the P-approximants graphically 
display a closer convergence to the numerical fluxes:  While being 
better than the Taylor approximation, they are still not as good as we 
would like.  However, we must remember that what is more important is how 
the phasing of the waveforms defined by the two approximants match with 
that defined by the exact flux, rather than a simple graphical agreement of 
the flux.  The graphical display is useful as it gives us a rough idea 
on any improvement in performance of a particular model.  
But as the noise of the interferometer weights the inner-product between 
templates, it may emphasize a particular range of frequency of the waveform
more than the other.  It may turn out that  the divergent region of a 
particular model is outside the detector bandwidth for most masses. In such cases
the approximant may work well in spite of the divergences or poles. 
We shall return to this question in Sec.~\ref{sec:results} 
when we consider the matching of waveforms.

\subsubsection{Retrograde Orbits, $q < 0$}\label{subsec:retromod}
We now focus on the GW flux from a test-mass counter-rotating with 
respect to the black hole.  We will again use a similar set of spin values, i.e. 
$q = \left\{-0.25,\, -0.50,\, -0.75,\, -0.95 \right\},$ as well as the Schwarzschild flux 
for comparative purposes.  In Fig.~\ref{fig:KTR}, left panels, we plot the 
Taylor approximant flux for retrograde orbits.   The solid lines in the 
figure correspond to the exact numerical fluxes with thicker lines
corresponding to smaller spin values $q = 0,\, -0.25,\, -0.50,\, 
-0.75,\,$ and $-0.95,$ respectively. The correspondingly thickened 
dashed lines represent the Taylor series approximation to the GW flux
with the same spin magnitude as the numerical flux. 
The right panels are the same except that in place of Taylor
approximant fluxes we plot the Inverse P-approximant
fluxes.

The first thing we notice is that just as in the case of prograde motion 
the ${\cal O}(x^{5})$ and ${\cal O}(x^{8})$ approximants are again highly 
divergent  although at ${\cal O}(x^{8})$, the convergence of the PN 
approximation improves as we go to larger negative spins. This is not 
entirely surprising:  As we go to larger 
counter-rotating spins the higher order Taylor expansion coefficients
become smaller and the non-linear general relativistic corrections
are not as important as in the case of prograde orbits. Moreover, 
the position of the LSO moves outwards from the BH as $q$ becomes
more negative.  At $q=-0.25$, the LSO is at $r_{LSO}\sim 6.8\,M$.  This 
corresponds to $x_{LSO} \sim 0.38.$  At $q=-0.95$ 
the LSO has moved outwards to $r_{LSO}\sim 9\,M$, corresponding 
to $x_{LSO} \sim 0.33$.  If we compare this to the equivalent 
spins in the prograde case we find that for $q=0.25$, $r_{LSO}\sim 5\,M$ and 
$x_{LSO} \sim 0.44$.  At $q=0.95$, $r_{LSO}\sim 2\,M$ and $x_{LSO} \sim 0.7$.  
In other words, retrograde orbits are weakly bounded orbits in comparison to
prograde orbits and experience smaller relativistic
corrections. Recalling that gravitational wave luminosity $\propto x^{10},$
retrograde flux for $q=-0.95$ at LSO is smaller by a factor $(0.70/0.38)^{10}
\simeq 500$ compared to prograde flux for $q=0.95$ at LSO.
Consequently, the PN approximation seems to work
well in the retrograde case. 

In the case of retrograde orbits P-approximants provide, 
at order ${\cal O}(x^{5}),$ only a marginal improvement 
over T-approximants.  This is in sharp contrast to the 
prograde case where the improvement was excellent.  At orders ${\cal O}(x^{6})$ and ${\cal O}(x^{8})$ are 
again quite good, with no poles at ${\cal O}(x^{7})$ for most spin values.  
While there seems to be a less obvious benefit in using the P-approximant flux in the retrograde case, 
it is again important to note that the P-approximants show the same general behaviour as the
exact flux in the retrograde case too and can therefore be expected to match the true general
relativistic signal more closely.

\subsection{Improving the performance of the Taylor- and P-approximants}\label{sec:bestfit}
In the last two Sections we noted how the P-approximant sequence seems to have a 
character similar to the sequence given by the numerical computations although the
two do not agree perfectly.  This suggests that if the 
value of the spin parameter used in the T- or P-approximant fluxes is varied relative to
the true value of the spin parameter in the exact flux, we should be able to 
obtain a better fit to the numerical fluxes.  While we will initially do this 
graphically, we remind the reader that this exercise is purely demonstrative.  
As the fluxes are Newton-normalized, it is difficult to say in what region of
the velocity we should attempt to make the best fit.  Added to this is the 
fact that when matching templates we should take the weighting of the PSD into 
account.  However, if the approximant fluxes agree graphically well with 
the numerical fluxes then it would mean that we should be able to match
the T- or P-approximant flux with the numerical result without introducing 
any new parameters.  To examine how best mis-matching spins might work we will 
focus on one of the post-Newtonian orders, say ${\cal O}(x^{6}),$ the order
at which the T-approximants work well and P-approximants have no spurious poles.

\begin{figure}[t]
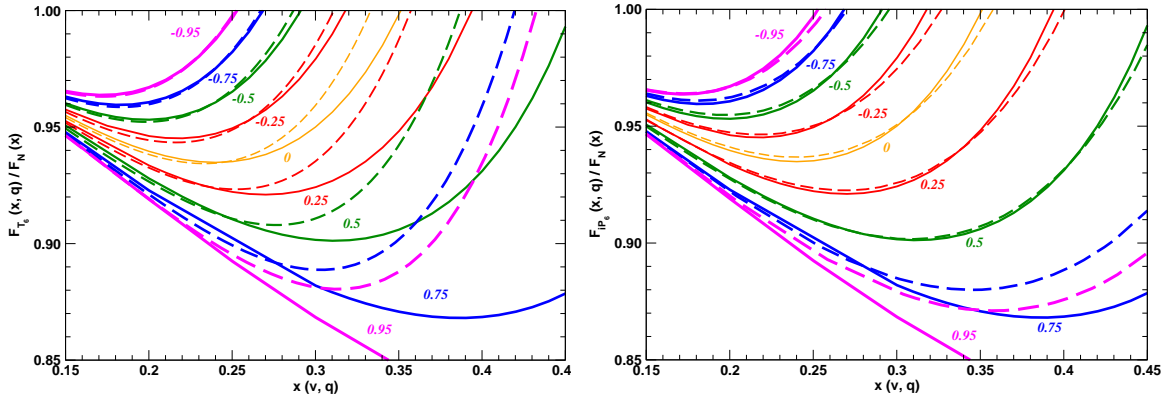

\begin{center}
\includegraphics[width=3in]{Tx6spinfit.eps}
\includegraphics[width=3in]{Px6spinfit.eps}
\caption{A best-fit for T-approximant (left) and P-approximant (right)
fluxes by varying the spin parameter 
$q$ at the ${\cal O}(x^{6})$ level for $q = -0.95$ (top most curve) through to $q = 0.95$ 
(bottom most curve).  Once again, the solid lines represent numerical fluxes and the 
broken lines represent the approximated fluxes.}
\label{fig:fitT}
\end{center}
\end{figure}

In Fig.~\ref{fig:fitT}, left panel, we have plotted the best-fit for the T-approximant flux 
for various values of the spin magnitude from $q = -0.95$ (top most curve) 
to $q = 0.95$ (bottom most curve).  For each spin magnitude we have fitted 
the T-approximant with the numerical result as best as possible
by varying the spin-magnitude of the approximant.  
In best fitting the numerical flux we incurred errors 
(defined as $\sigma_q \equiv 1 - q_{\rm fitted}/q_{\rm exact}$) in the
spin magnitude of the approximant relative to the exact case. These are listed
in Table \ref{table:bestfit}.  Although the errors 
decrease as we approach the extreme prograde flux, we can see from the figure 
that the fit is not very good for the Taylor approximant for $q\ge 0.$
Schwarzschild case.  
\begin{table}
\caption{The ``best-fit'' value of the spin parameter, $q$, for T and 
P-approximant fluxes for $q = -0.95$ to $q = 0.95$.
For the Schwarzschild case we have an absolute error of 0.11 in the case of 
T-approximant and $-0.08$ in the case of P-approximant. }
\begin{tabular}{c|ccccccccc}
\hline
\hline
Spin Magnitude & $-0.95$   & $-0.75$   & $-0.50$ & $-0.25$   & $0.25$  & $0.50$  & $0.75$ & $0.95$  \\ 
\hline
$q_T$          & $-0.90 $  & $-0.68$   & $-0.42$  & $-0.13$   & $ 0.35$ & $0.62$ & $ 0.89$ &  $0.99$\\ 
$q_P$          & $-0.99$   & $-0.81$   & $-0.60$  & $-0.33$   & $ 0.21$ & $0.55$ & $ 0.90$  & $0.99$\\
\hline
$\sigma_{q_T}$ & $5.3\%$  & $9.3\%$   & $16\%$  & $48\%$    & $40\%$  & $24\%$  & $19\%$ & $4.2\%$   \\
$\sigma_{q_P}$ & $4.2\%$   & $8.0\%$     & $20\%$  & $32\%$    & $16\%$  & $10\%$  & $20\%$ & $4.2\%$ \\
\hline
\hline
\label{table:bestfit}
\end{tabular}
\end{table}

We have plotted the best-fit results for the P-approximant
in right panel of Fig.~\ref{fig:fitT}.  We can observe that while the fit is not as
good as the PN flux in the extreme retrograde case, it gets superior as we 
move towards the extreme prograde flux.  In fact, the P-approximant best-fits work 
extremely well up to a spin of $q=0.5.$ It also works well at $q=0.75,$ but
only at low values of the parameter $x.$ 
For $q=0.95$ we did not obtain a good fit without introducing a new parameter.
It is clear that there is an advantage in using the P-approximant flux. However, we will 
have to wait until we calculate the fitting factors between templates to see just how 
important this advantage is.

Let us conclude this Section by noting that when approximate fluxes, both T- and P-approximants,
best fit the exact numerical flux, errors in relative spin magnitude are large at low 
values of spins, i.e.  several 10's of percents for $|q| \le 0.50,$ while the errors 
are low at large spin magnitudes, i.e. a few percents for $|q| \ge 0.75.$ We will see 
that this will exactly be the case even when we compute the best overlap of the 
approximate signal with the exact signal.


\section{Effectualness and Faithfulness of T- and P-Approximants}\label{sec:results}
In Sec.~\ref{sec:waveform} we have seen that the phase of the GW depends
on the evolution of the binary's binding  
energy and gravitational wave flux functions.  As the energy is 
known exactly for a test-particle in circular orbit about a
Kerr black hole it is important to compute the flux to sufficiently
high accuracy so that we can match the phasing of the exact waves
to a good precision.  Investigations of the PN approximation 
showed that while T-approximants have bad convergence properties
P-approximants improved the convergence of the flux.
We also found that we could use the spin parameter $q$ as a free 
parameter in order to obtain a better matching of the flux.  
By best fitting the spin parameter, we were able to graphically 
achieve a closer fit to the numerical fluxes.   Equipped with the new,
improved, P-approximant we now move onto explore
how well a waveform based on it matches the phasing of the exact waves. 
We shall address the performance of the approximants in extracting
the exact waveform in two ways: The {\it effectualness} of the templates
measured in terms of the maximum overlap they can achieve with the exact waveform
when the parameters of the approximant are varied in order to achieve
a good match. The {\it faithfulness} of the approximant templates 
measured in terms of the systematic errors in the estimation 
of parameters while detecting exact waveforms.

\subsection{Overlaps and fitting factor}
We can employ the results of matched filtering, and its geometrical
interpretation, to assess how well our approximant waveforms match
the exact waveform. The geometrical theory of signal analysis \cite{BSD1,BJO}
can be summarized as follows: The set of all detector outputs,
each lasting for a duration $T,$ and sampled at a rate $f_s,$ and 
consisting of $N=f_s T$ samples, can be thought of as a linear
vector space. Parametrized signals, such as a binary inspiral waveform
that depends on the two masses of the component stars and their spins,
are also vectors in the vector space of all detector outputs. However,
the set of all waveforms do not form a vector
space although they do form a manifold with the parameters serving
as coordinators. 

Matched filtering technique, which is the most
effective technique to capture signals of known phase evolution,
can be used to define a metric on such a manifold. The metric
and the associated scalar product allows us to compute the distance
between any two signal vectors that are parametrized in 
the same way. The scalar product is the same as that introduced in 
Eq.~(\ref{eq:scalarprod}). If we have two different families of 
waveforms, for example T- and P-approximants, parametrized similarly
then each family lives on a distinct manifold. The distance between
two vectors with exactly the same parameters but belonging to
different families is, in general, not zero. Although the scalar product
can be computed between any two vectors in the space it is only in 
the case of two similarly parameterized signal vectors does the scalar product
represent the distance between the vectors. This is  because the coordinates,
that is the parameters, don't have any meaning for vectors that don't
live on the manifold. By calculating the 
scalar product between two normalized vectors, that is vectors whose
scalar product with themselves is unity, we can see how well they match. 
For two normalized waveforms, or signal vectors, the scalar product returns the 
cosine of the angle between them and is is normally referred to as 
the {\it overlap,} denoted by $O$.  Given two waveforms $h$ and $g,$ not 
necessarily belonging to the same family of approximants, their 
overlap is defined as
\begin{equation}
{\mathit O} \equiv 
\frac{\left<h\left|g\right>\right.}{\sqrt{\left<h\left|h\right>\right.\left<g
\left|g\right>\right.}},
\end{equation}
where, the inner-product between two real functions $h(t)$ and $g(t)$ is 
defined by Eq.~(\ref{eq:scalarprod}).  For detection of signals
what is more important 
is the {\it fitting factor} $FF:$  As each template is a function 
of the intrinsic parameters $\lambda^{\mu}$, the fitting factor is defined as 
the maximum overlap obtained by varying the parameters of the template (or the
approximate waveform) relative to the exact waveform:
\begin{equation}
FF = \max_{\lambda^{\mu}} {\mathit O}\left(\lambda^{\mu}\right).
\end{equation}


\begin{figure}[t]
\begin{center}
\includegraphics[width=3in]{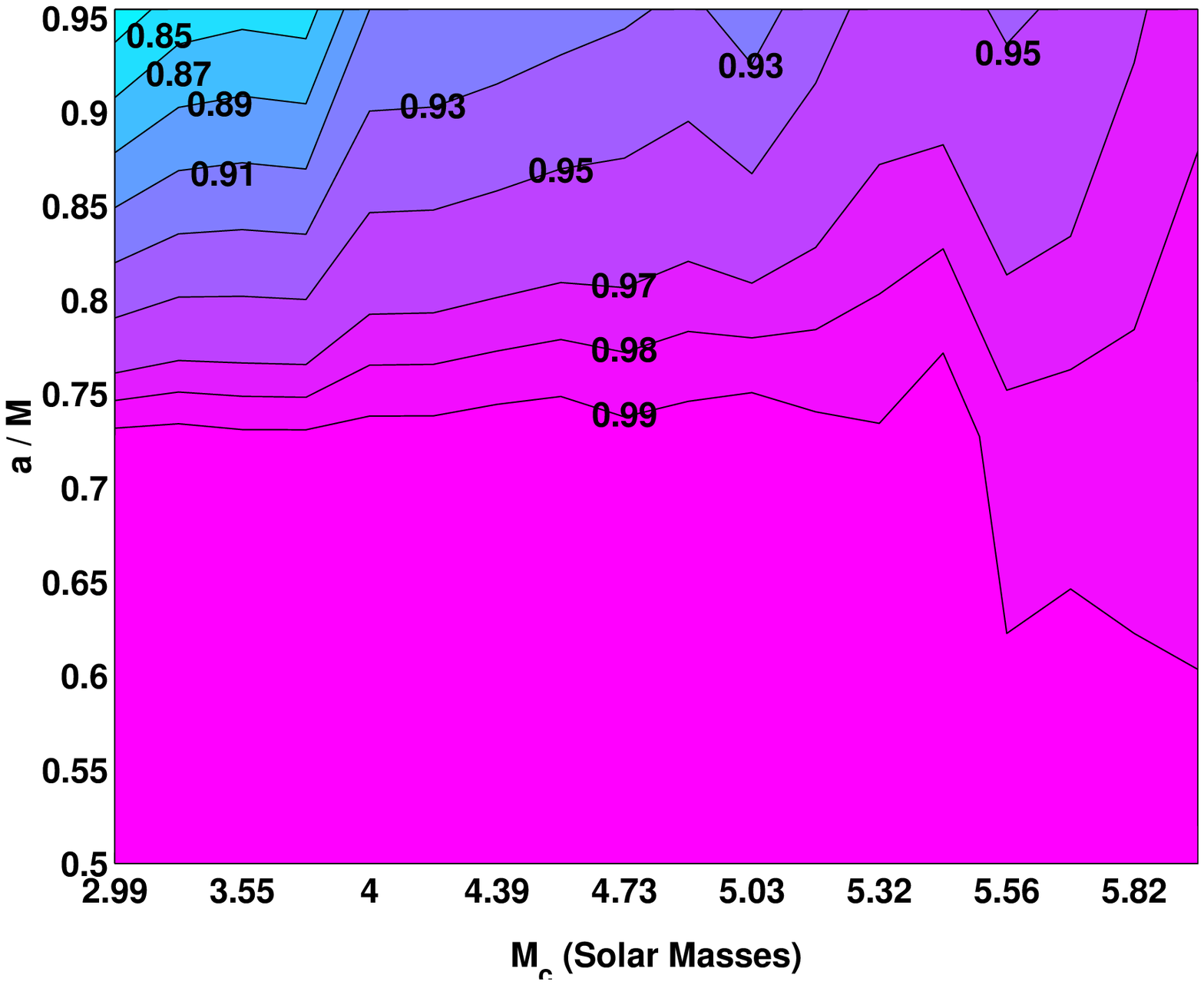}
\includegraphics[width=3in]{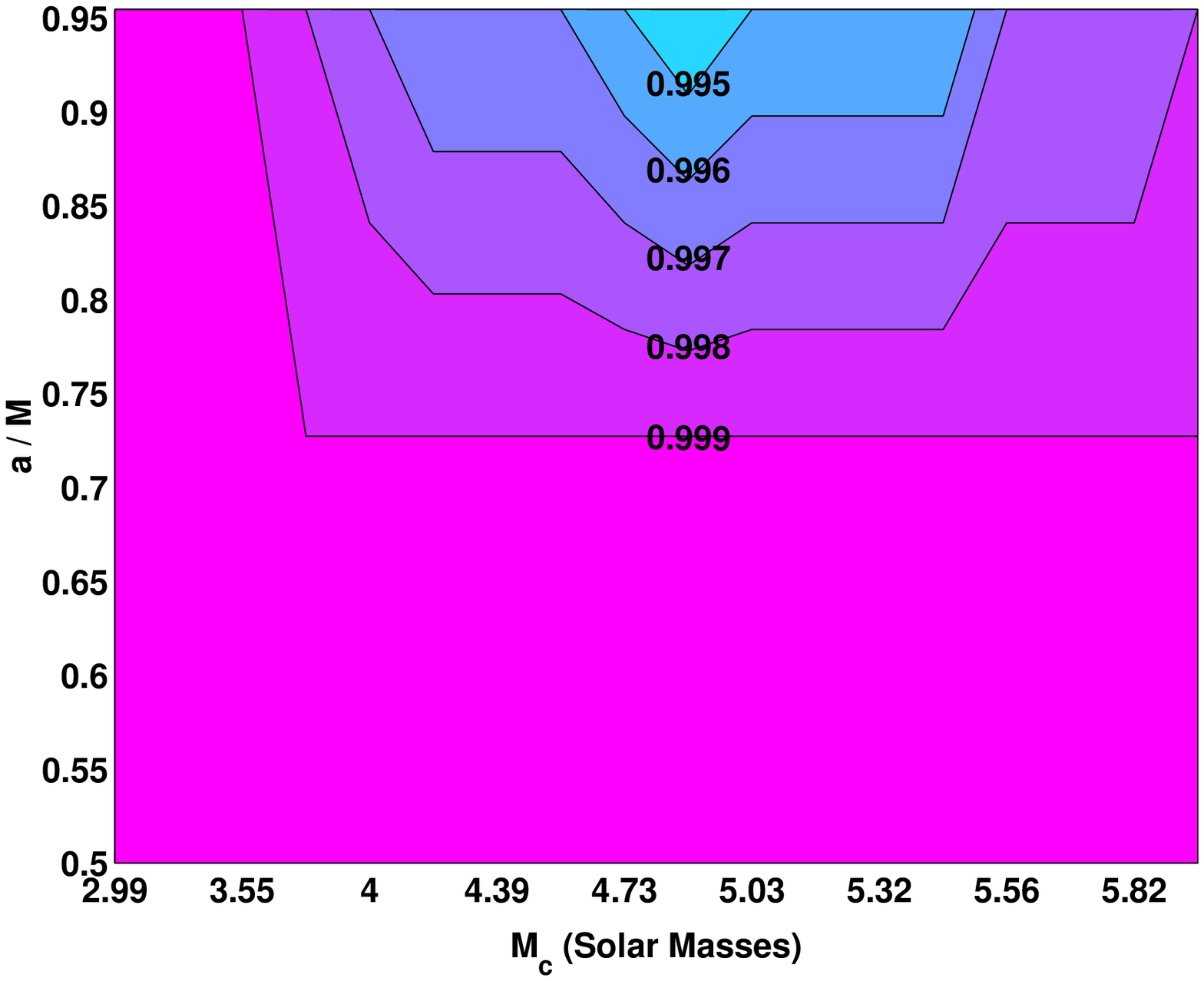}
\caption{The maximized prograde overlaps for T-approximant (left) and
P-approximant (right) templates at the 
$x^{8}$ approximation.  Each system consists of a $1.4\,M_{\odot}$ NS 
inspiralling into a central BH of mass ranging from 10-50 $M_{\odot}$.  The 
figure covers prograde Kerr from $q = 0.5$ onwards, as fitting factors of $> 0.99$ are achieved for all lower spins.}
\label{fig:TO}
\end{center}
\end{figure}  

If two waveforms are a perfect match then their overlap is unity.  As the 
waveforms begin to differ, their overlap differs from 1 and the value of
the overlap is a measure of how similar the two waveforms are and how useful
one waveform is in extracting the other waveform buried in noise.  Now the GW from
a binary is determined by a number of parameters: The two masses, spins, eccentricity
and the orientation of the angular momentum at some initial time. These parameters
essentially determine the dynamics of the system and are called {\it intrinsic}
parameters \cite{BJO}. The observation of such a system introduces five other parameters:
These are the instant $t_0$ of coalescence (or the time at which the orbital frequency
reaches a certain value) and the phase $\phi_0$ of the system at that instant, the angular
position of the system in the sky, the distance of the binary (or, equivalently, the
amplitude of gravitational waves at Earth) and the polarisation angle (or, equivalently,
the ratio of $h_+$ and $h_\times$). These are called the {\it extrinsic} parameters \cite{BJO}
and do not play any role in the dynamics of the system. For a signal that lasts
only for a few mins, as would be the case for systems expected to be observed
in a ground-based interferometer, the direction and the polarisation angle cannot be
measured in a single interferometer and the distance is only an amplitude parameter.
Thus the templates are only needed for parameters $t_0$ and $\phi_0.$ 
In the case of circular equatorial orbits of a test mass around a central
black hole, the wave is parameterized by the $(t_{0},\, \Phi_{0},\, M,\, \eta,\, q).$ 

In order to improve the detection 
probability, we would like to maximize over as many of the parameters as 
possible.  Maximization over $t_{0}$ is achieved by simply 
computing the correlation of the template with the data in the frequency domain
followed by the inverse Fourier transform. This yields the correlation of the
signal with the data for all time-lags. However, one has to worry about circulation
correlation effects, and the consequent corruption of the correlation when the
template gets split at the boundaries of the sample sets, but these are easily handled
by padding the template with sufficiently large number of zeroes before performing
the Fourier transform.

It was pointed out by Schutz \cite{Schutz2,DhurSath} that the
maximum of the correlation $C$ of data with a template
over $\Phi_{0}$ can be computed using 
just two templates -- an {\it in-phase} and a {\it quadrature-phase}
template.  In other words, generate two orthonormalized 
templates with phase $\Phi_{0} = 0$ and $\Phi_{0}= \pi / 2$.  
The maximized overlap is then given by 
\begin{equation}
\max_{\Phi_0} C = \sqrt{C_{0}^{2} +  C_{\pi/2}^{2}},
\end{equation}
where $C = \left<h^A \left(\Phi_{0}\right)\left|\right . h^{\rm X}\right>,$ 
$C_{0} = \left<h^A \left(\Phi_{0} = 0\right)\left|\right .h^{\rm X}\right>$ 
and $C_{\pi / 2} = \left<h^A \left(\Phi_{0} = \pi / 2\right)\left| \right . h^{\rm X}\right>$.  Here, 
$h^{\rm X}$ denotes some normalized ``exact'' waveform.  However, numerically, this 
procedure does not fully orthogonalize the two templates.  
\begin{figure}[t]
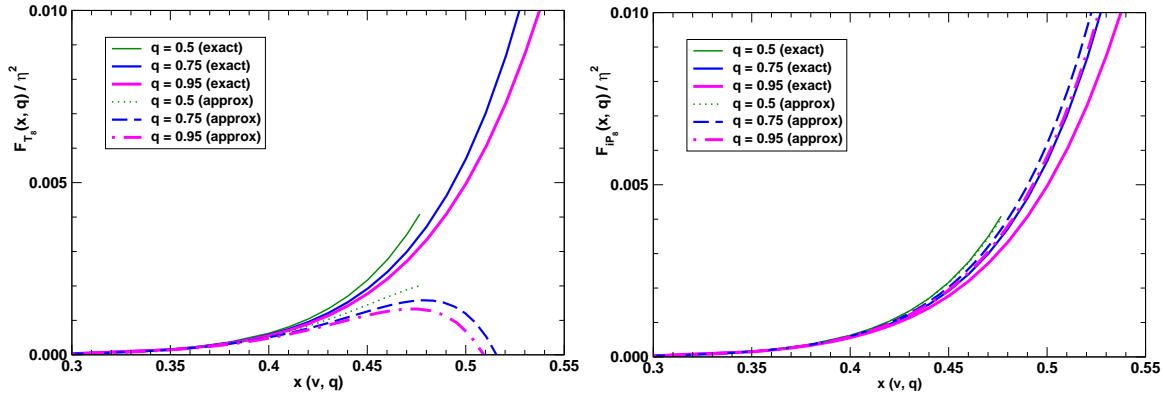

\begin{center}
\includegraphics[width=3in]{T8Kerrfluxfunc.eps}
\includegraphics[width=3in]{P8Kerrfluxfunc.eps}
\caption{The 4-PN T-approximant (left) and
P-approximant (right) analytical flux function for spins of $q = 0.5, 0.75$ 
and $0.95$ against the numerical fluxes for the same spin values. 
We can see that, beyond $q = 0.5$,  the T-approximant flux goes to 
zero before the LSO is reached.  This happens sooner, the higher 
the spin value we go to.  This is not a problem for the P-approximant flux.}
\label{fig:x8ff}
\end{center}
\end{figure}  
For this reason we specifically orthonormalize the two waveforms.  
Using two un-normalized waveforms $\tilde{h}_{0} = 
\tilde{h}\left(\Phi_{0} = 0\right)$ and $\tilde{h}_{\pi /2} = 
\tilde{h}\left(\Phi_{0} = \pi /2\right)$, we generate two orthonormalized 
waveforms according to 
\begin{equation}
\tilde{H}_{0} = \frac{\tilde{h}_{0}}{|\tilde{h}_{0}|},\ \ \ \ 
\tilde{H}_{\pi /2} = \frac{\tilde{h}_{\pi /2} - \left<h_{\pi 
/2}\left|H_{0}\right>\right.\tilde{H}_{0}}{| \tilde{h}_{\pi /2} - \left<h_{\pi 
/2}\left|H_{0}\right>\right.\tilde{H}_{0}|}.
\end{equation}
Alternatively, one could define the quadrature-phase template using
$\tilde{H}_{\pi/2} \equiv i \tilde{H}_{0},$ which
is explicitly orthogonal to the in-phase template.
The (square of the) maximum of the overlap over $\Phi_{0}$ is given by the
sum-of-squares of the correlation with the in-phase and quadrature-phase templates:
\begin{equation}
\max_{\Phi_0}\left< C \right> = \sqrt{\left< 
H_{0}\left|h^{\rm X}\right.\right>^{2} + \left<  
H_{\pi/2}\left|h^{\rm X}\right.\right>^{2}}.
\end{equation}
Thus, the correlation can be easily maximized over 
the unknown time-of-arrival and the constant phase of the templates.
However, we can also maximize over all other parameters 
using a grid of templates.

In this paper, as we are working in the test-mass approximation, we assume that 
our system is composed of objects with a small mass ratio. For concreteness
we assume that the system comprises a $1.4 M_{\odot}$ NS inspiralling into a central 
BH of varying spin magnitude and mass.  Beginning with a central BH of $10 M_{\odot}$, we 
work our way upward to a $50 M_{\odot}$ BH.  In terms of the symmetric mass 
ratio, this gives us a range of test systems from $\eta = 0.027$ for the 
heaviest binary in our list [i.e. $(50$-$1.4)M_\odot$ system] to 
$\eta = 0.11$ [i.e., $(10$-$1.4)M_\odot$ system].  
We will also look at the limiting case of a 
$10$-$10 M_{\odot}$ equal-mass system.  Strictly speaking, 
the formulas for energy and flux functions used
in this study are not applicable to the comparable mass case since 
we have neglected the finite mass correction terms in these quantities.
However, the results of such a study should give us an indication
of how strong are the relativistic corrections, as opposed to finite-mass
corrections, in the case of comparable masses.
In all cases we vary the spin magnitude of the central BH 
from $q = -0.95$ to $0.95$.  

In the next two sub-sections, we will focus respectively on prograde and 
retrograde systems.  As well as presenting the results for the fitting factors achieved, we 
will also focus on the problem of parameter extraction using templates constructed
from the Taylor- and P-approximants.  
Since the main focus in this study is test-mass approximation we shall be
interested in only one of the mass parameters and consider
errors in estimating the chirp-mass ${\mathcal M} = m\eta^{3/5}$ in addition to
the spin magnitude of the central BH.  In view of economy we shall 
only present the results for the highest PN order available, namely
${\cal O}(x^{8})$ order. In all cases our fiducial {\it exact} 
signal $h^{\rm X}$ will be that obtained by using the exact expression for the energy in 
Eq.~(\ref{eq:energy}) and the exact numerical fluxes generated using
black hole perturbation theory \cite{Shibata},
and the template will be the approximate waveform constructed using the
exact expression for the energy, as before, and an approximate expression for
the flux, either the T-approximant flux, the corresponding template
denoted $h_{\rm T},$ or the P-approximant flux, the corresponding template
denoted $h_{\rm P}.$


\begin{figure}
\begin{center}
\includegraphics[width=3in]{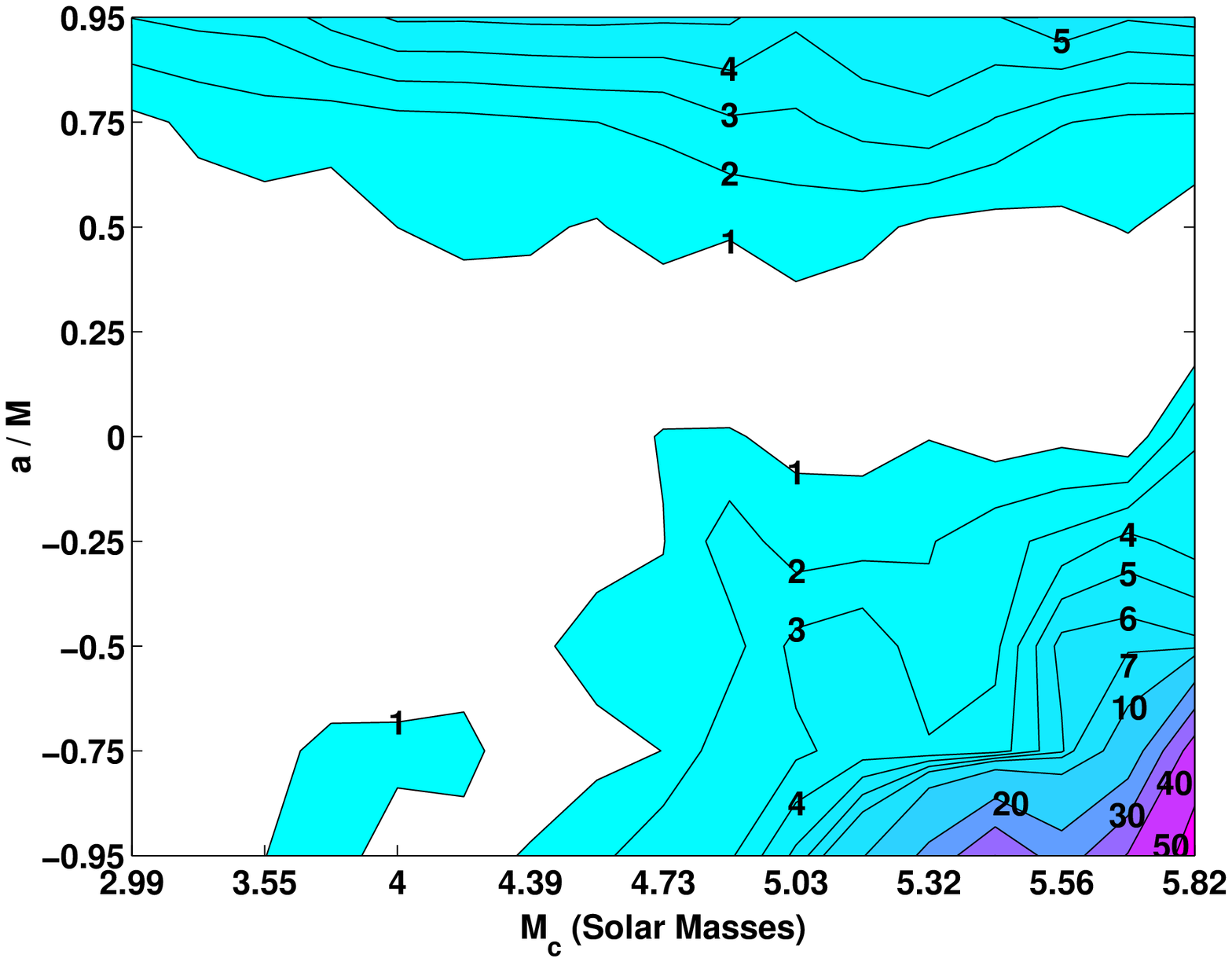}
\includegraphics[width=3in]{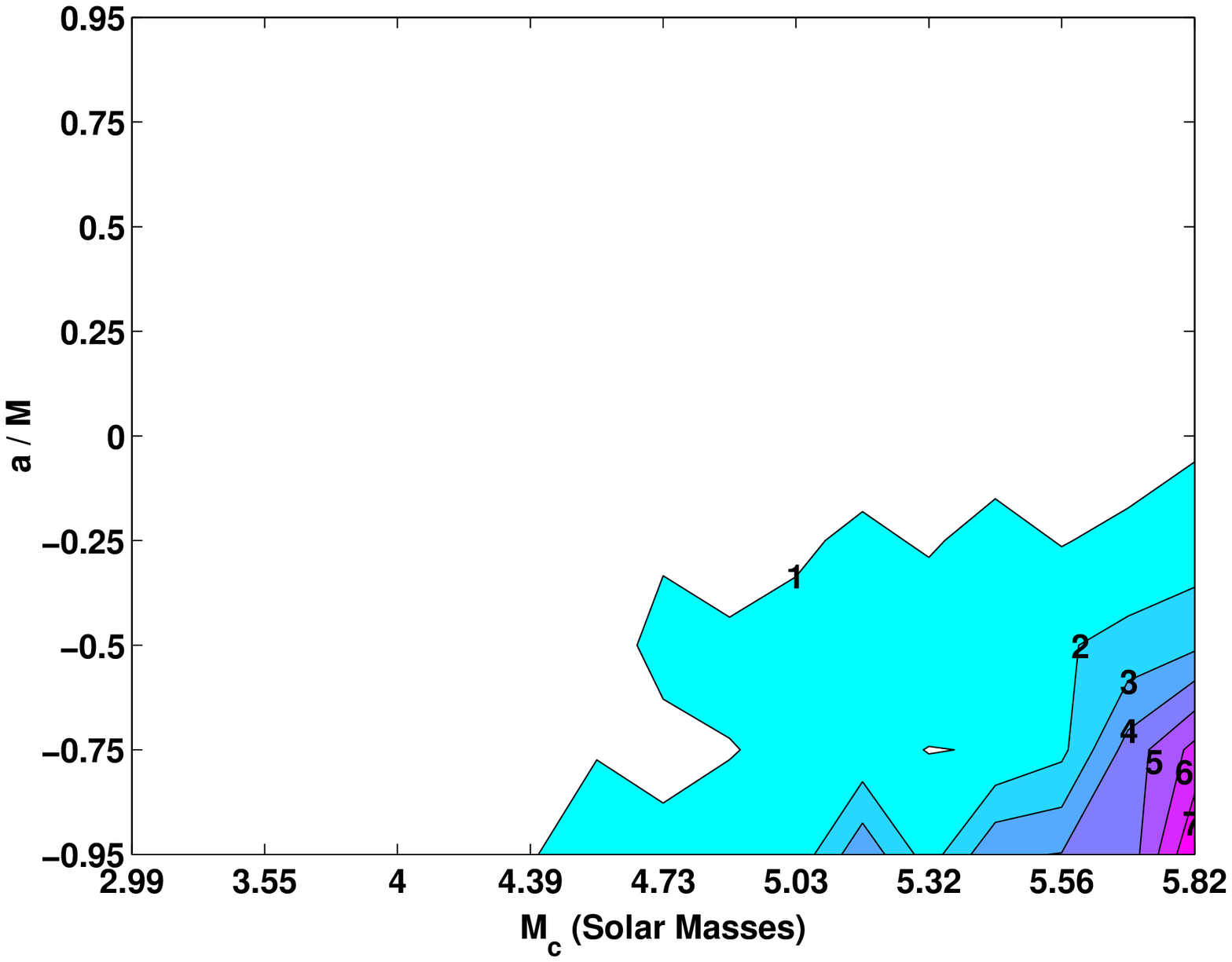}
\caption{The percentage error in the estimation of the chirp-mass, ${\mathcal 
M}_{c}$, for T-approximant (left) and P-approximant (right) templates at 
the $x^{8}$ approximation.   The values 
of ${\mathcal M}_{c}$ correspond to a $1.4\,M_{\odot}$ NS inspiralling into a 
central BH of mass ranging from 10-50 $M_{\odot}$.  The figure covers 
retrograde Kerr, Schwarzschild and prograde Kerr systems.}
\label{fig:TCME}
\end{center}
\end{figure}  

\subsection{Prograde Orbits -- Effectualness}
In the first instance, we maximized the overlap only
over $t_{0}$ and $\Phi_{0}$, while holding all other parameters 
of the signal and the templates the same.  The overlaps so obtained were not
good enough.  In the case of T-approximant templates the overlaps at $q = 0.25$ 
were in the range $0.6 \leq FF \leq 0.8$, depending on the PN order.  By $q 
= 0.95$ they had fallen to $0.3 \leq FF \leq 0.5$.  The P-approximants 
actually fared much worse in this case.  At $q = 0.25$ the fitting factors 
lay between $0.25 \leq FF \leq 0.65$, while at $q = 0.95$, none of the 
fitting factors achieved was $> 0.4$.

\begin{figure}
\begin{center}
\includegraphics[width=4in]{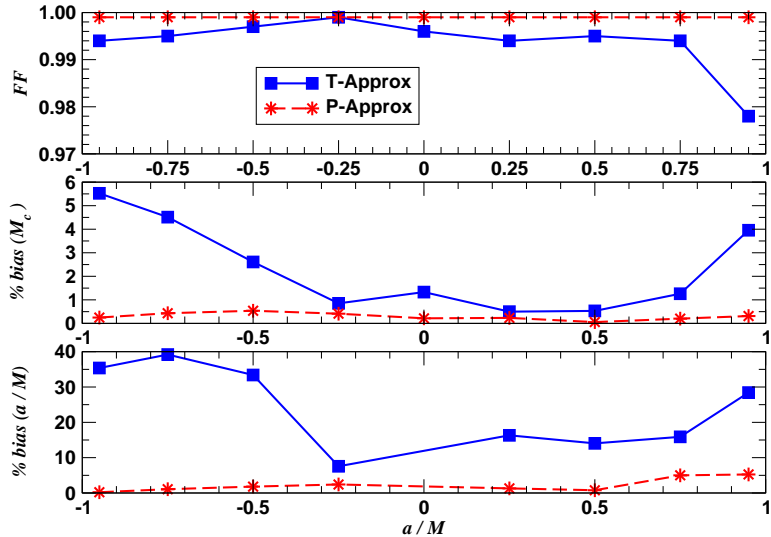}
\caption{The fitting factors for a 10-10 $M_{\odot}$ binary - without the 
finite mass correction terms - for T and P-approximant templates (top).  The 
percentage error in the estimation of the chirp-mass for T and P-approximant 
templates (middle). The percentage error in the estimation of the spin of the 
central black hole for T and P-approximant templates (bottom). }
\label{fig:eqmasspro}
\end{center}
\end{figure}

It is clear that maximizing over extrinsic parameters is not adequate:
When we maximize over all other parameters, overlaps improve significantly
as compared to the unmaximized overlaps. 
For T-approximants -- Fig.~\ref{fig:TO}, left panel -- maximizing over all parameters gives 
fitting factors of ${FF \geq 0.98}$ at all mass ranges for the test-mass systems 
up to a spin of $q=0.75$.  Between 0.75 and $q=0.95$, the T-approximant 
templates begin to perform badly and the fitting factors drop to ${FF \sim 
0.82}$.  For the equal-mass case -- Fig.~\ref{fig:eqmasspro} --  the templates 
once again achieve fitting factors of ${FF \geq 0.99}$ up to a spin of 
$q=0.75$, but fall off at higher spin magnitudes achieving a fitting factor of 
$\sim 0.98$ at $q = 0.95$.  We should point out that these results do not 
properly convey just how bad the 4-PN T-approximant template actually performs.  
In the left hand panel of Fig.\ref{fig:x8ff}, we plot the PN approximation 
for the flux function against the numerical fluxes at spins of $q = 0.5, 0.75$ 
and $0.95$.  We can see that the flux function at $q = 0.75$ and $0.95$ become 
negative long before the LSO is reached.  This means that as we go to higher and 
higher spins, we can model less and less of the waveform.  
For $q \le 0.6$ we can model the waveform up to the LSO.  However, for $q>0.6$
this we have to stop the waveform generation at a cutoff velocity of 
$x_{cut} \approx 0.5$.  At $q = 0.75$ we model approximately 
$80\%$ of the waveform, while at $q = 0.95$, this reduces to $44\%$.  
Thus, T-approximants have to be prematurely terminated well before
the LSO is reached.  In the process of maximization of the overlap we 
systematically find that exact waveforms corresponding to a given total 
mass are maximized by T-approximant templates of significantly 
{\it smaller} masses.  The reason for this is that smaller mass 
templates will have a longer duration (and therefore make up for 
the loss in overlap due to premature termination of the waveform) and 
thereby achieve a better overlap with the exact waveform.

Therefore, the fitting factors beyond $q = 0.75$ completely overestimate 
the performance of the T-approximant template.

If we now focus on the right hand panel of Fig.\ref{fig:x8ff}, 
we see the true power of the P-approximant templates.  The 4-PN 
template suffers none of the divergences that effect the T-approximants.  
We therefore generate all templates up to the LSO or 2 kHz, whichever 
is reached first.  We can see from the right panel of Fig.~\ref{fig:TO} 
that the P-approximant templates achieve fitting factors of $> 0.99$ 
at all spin values.  For the equal-mass case 
-- Fig.~\ref{fig:eqmasspro} -- the P-approximant templates achieve fitting 
factors of ${FF \geq 0.995}$ at all mass and spin levels.  This demonstrates 
that in the case of prograde orbits, the P-approximant templates are clearly 
more robust, even at high spin magnitudes of $q=0.95.$

\subsection{Prograde Orbits -- Faithfulness}
Gravitational wave observations of the inspiral signal are expected to 
result in very precise measurements of the signal parameters. Indeed, it
has been shown \cite{CF94} that the chirp mass can be measured to a relative 
accuracy of $0.02\%$ and $0.16\%$ respectively for systems comprising 
$(10,\,1.4) M_\odot$ and $(10,\,10)M_\odot$ objects when the correlations
between the spin magnitude and masses is neglected and
to an accuracy of $0.19\%$ and $1.42\%$ for the same systems when
the correlations are included.  Such accurate estimation
of parameters will aid in high precision tests of general relativity.
When we use approximate templates the question naturally arises if the
{\it bias} in the estimation of parameters
\footnote{While using approximate templates to detect a truly general
relativistic signals a bias in the estimation of parameters is inevitable.
This is because, as noted while studying the effectualness of templates,
the template shape is not identical to that of the signal when the 
intrinsic parameters are matched. Consequently, higher overlaps can be
achieved by mismatching the template parameters  relative to the signal.
Since matched filtering isolates the template that obtains the best signal-to-noise
ratio, there will be a  bias in the estimation of parameter.}
is smaller than the
accuracy with which the parameters can be determined. This is what we shall
consider next.

While the T-approximant templates performed well up to $q=0.75$, a 
clear indication of how good they truly are comes from the bias in parameter 
estimation.  This tests the faithfulness of the templates.  It should be 
pointed out at the outset that there is a large covariance between the chirp
mass and the spin magnitude and a small error in the estimation of 
the chirp mass is compensated by  a large error in the estimation of the 
spin magnitude.  In Fig.~\ref{fig:TCME} we have 
plotted the percentage bias in the estimation of the chirp-mass for both T-
(left panel) and P-approximant (right panel) templates.  From
Fig.~\ref{fig:TCME}, left panel, it is clear that for T-approximants
the bias in ${\mathcal M}$ varies between 0 -- 5$\%.$  
Comparing the left and right panels of Fig.~\ref{fig:TCME}
we find that the P-approximant templates (right panels) are more faithful with the 
percentage bias being less than $1\%$ in general.  

\begin{figure}
\begin{center}
\includegraphics[width=3in]{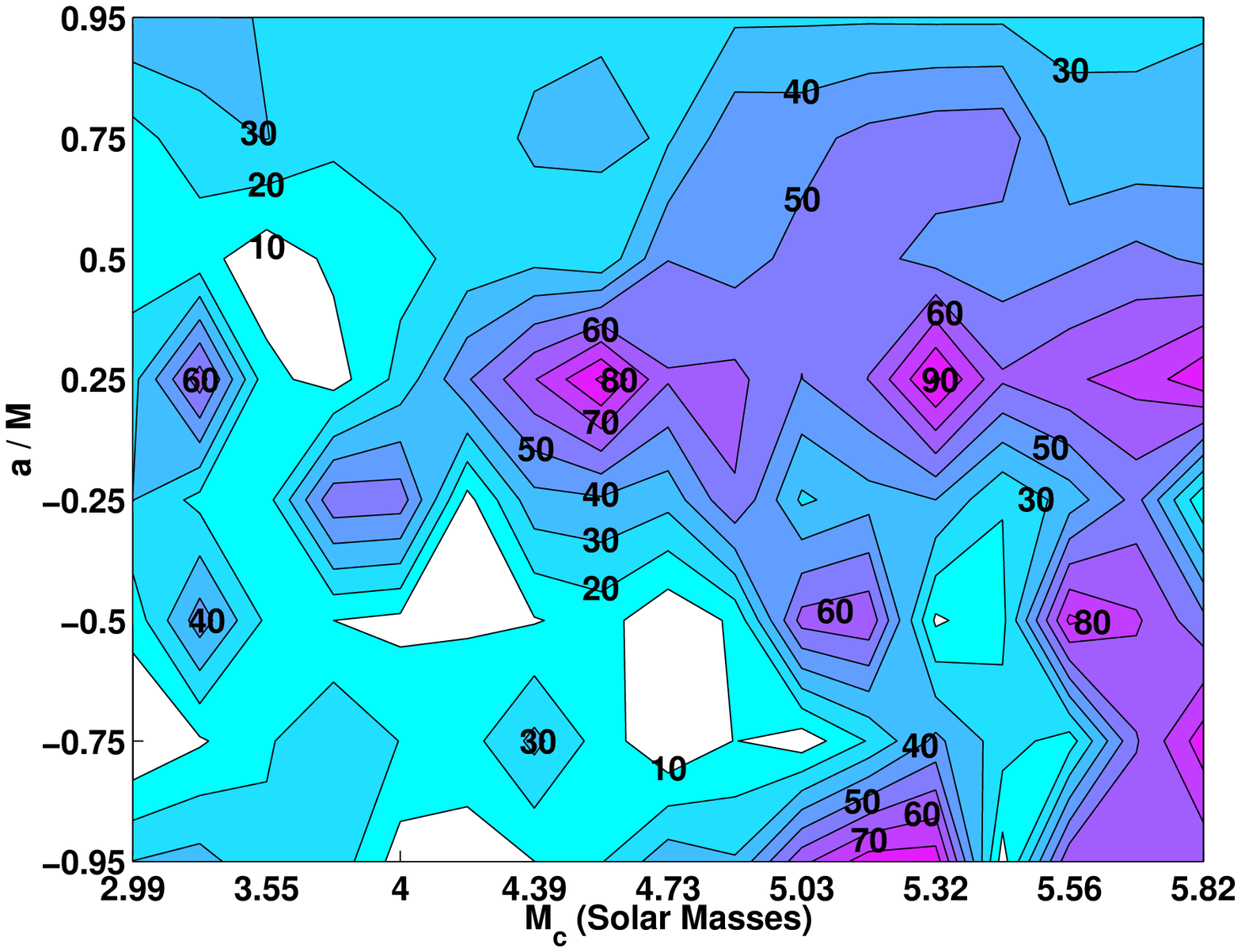}
\includegraphics[width=3in]{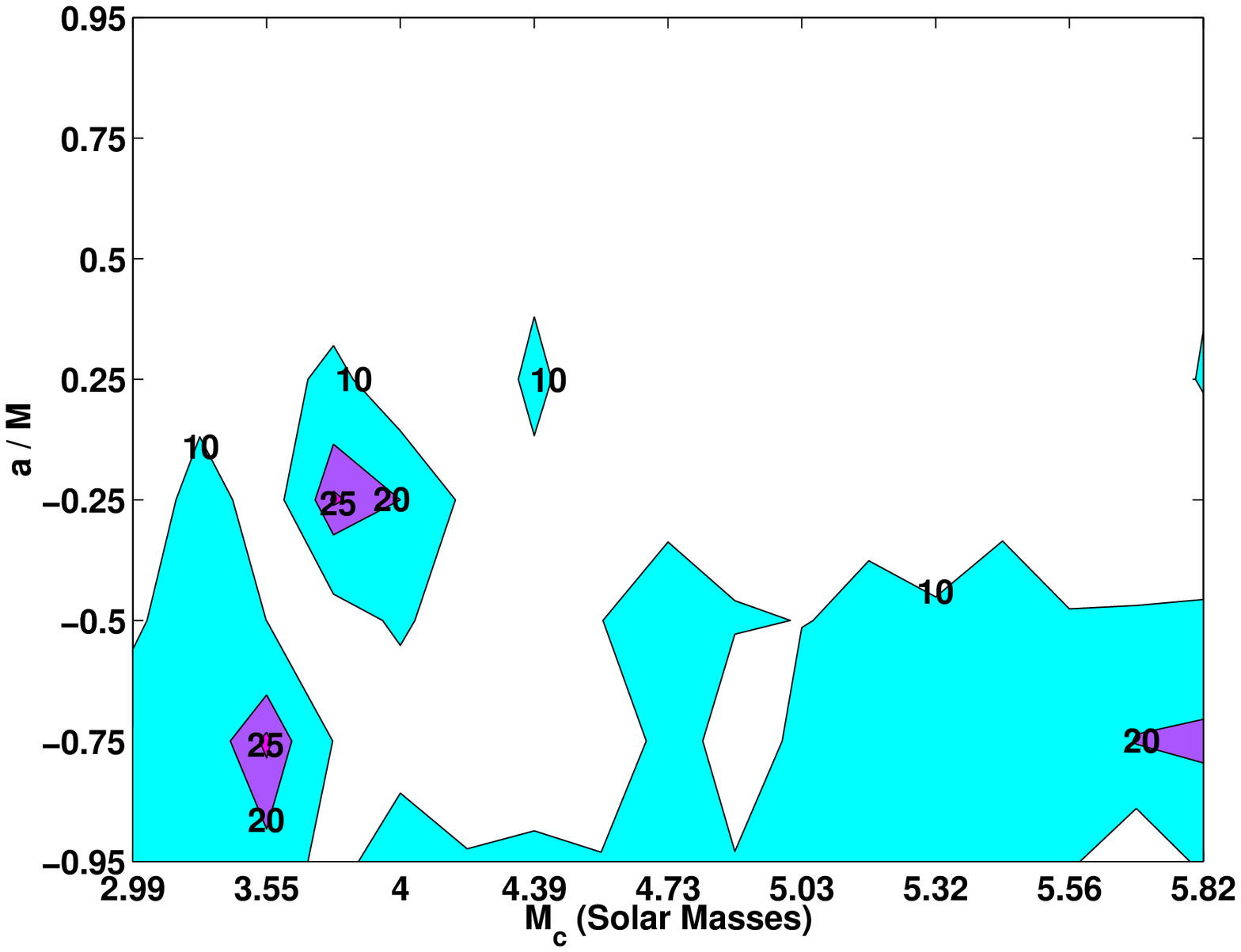}
\caption{The percentage error in the estimation of the spin parameter, $q$, for 
T-approximant (left) and P-approximant (right) 
templates at the $x^{8}$ approximation. The values of ${\mathcal 
M}_{c}$ correspond to a $1.4\,M_{\odot}$ NS inspiralling into a central BH of 
mass ranging from 10-50 $M_{\odot}$.  The figure covers retrograde and prograde 
Kerr systems. }
\label{fig:TAE}
\end{center}
\end{figure}

In Fig.~\ref{fig:TAE} we present the percentage bias in the estimation 
of the spin magnitude of the central BH.  We find that for both 
approximants we have to endure a large bias in $q$.  Once again the 
P-approximant templates are more faithful.  For T-approximants the bias 
varies between 5 and 90$\%$, while for P-approximants it lies between 2 and 10$\%$.  
From the middle and bottom Sections of Fig.~\ref{fig:eqmasspro}, it is clear that
even in the equal mass case the bias in the estimation of both parameters 
is greater for T-approximant templates than P-approximants.

It is therefore clear that in the case of prograde orbits, P-approximant 
templates must be used in any detection strategy.  Not only do they achieve 
higher fitting factors at all spin and mass ranges compared to the 
T-approximant templates, but they are also more faithful.  Let us finally 
note that the bias in the estimation of spin is larger at low spin values,
which is exactly what we found by best fitting the approximate fluxes with
the exact fluxes by using $q$ as a free parameter.


\subsection{Retrograde Orbits -- Effectualness}
In the case of retrograde motion we carried out the same range of tests as in 
the prograde case.  We found that in the case of maximization over $t_{0}$ and 
$\Phi_{0}$ the overlaps are not as bad. It 
seems that all templates perform reasonably well for all spin values.  This is 
probably due to the fact that the PN coefficients at higher orders are smaller when
$q<0$ than when $q>0.$ Additionally, retrograde orbits are less bound and, therefore, 
less relativistic, and again higher order PN corrections do not play as significant
a role for retrograde orbits as they do for prograde orbits.  At a 
spin value of $q = -0.25$, the T-approximant templates achieved fitting factors in 
the range $0.67 \leq FF \leq 0.81$.  This falls off to a mean value of $\sim 
0.54$ at $q = -0.95$.  The P-approximant templates also perform better in 
the retrograde case achieving fitting factors in the range $0.47 \leq FF \leq 
0.80$ at $q = -0.25$, and between $0.54 \leq FF \leq 0.84$ at $q = -0.95.$

In addition to extrinsic parameters when we maximize over all 
the intrinsic parameters the fitting factors improve.  In fact, in the case of retrograde 
orbits both templates achieve fitting factors of $FF \geq 0.99$ regardless of 
the spin magnitude and the chirp mass. There is no surprise here: 
The retrograde waveforms are still well within the adiabatic regime and 
are, therefore, modelled well by both templates.  From a purely detection point of 
view, unlike the prograde case, there is no obvious benefit from employing 
P-approximant templates.  For the equal mass system too
there is not much difference in the fitting factors of the two
families of  templates with the exact signal.  However, the 
fitting factor for T-approximants is well below than the
corresponding P-approximant template for spin magnitudes $q>0.5.$

\subsection{Retrograde Orbits -- Faithfulness}
The benefit of using P-approximant templates for retrograde motion is only 
observed when we consider parameter extraction.  Referring to the bottom
portions of the two panels in Fig.~\ref{fig:TCME}, we note that the 
T-approximants perform well at all spins for $3.0 \geq {\mathcal M} \leq 
4.5$, with a bias of less than $1\%$ in the estimation of ${\mathcal M}$.  
Beyond this, there is in general a bias of $>$ 2$\%$.  As we approach the 
extreme retrograde case the bias rises to as much as 55$\%$.  
The P-approximants perform in a similar manner.  The bias in the 
region $3 \geq {\mathcal M} \leq 4.5$ is again in general less than $1\%$.  The 
error does again increase as we head towards the extreme test-mass range, but 
in this case it reaches a maximum value of $8\%$ as opposed to the $55\%$ seen
in the case of  the T-approximants.

The pattern observed in the retrograde case is similar to the 
prograde case with the errors in the 
T-approximants greater than those in the P-approximants.  The error in 
estimating $q$ reaches a maximum of $80\%$ for T-approximants and $25\%$ for 
P-approximants.  There is a slight difference, however, between the two cases:  
We observe that there is an asymmetry in the distribution of contour lines.  
The T-approximant templates suffer a higher error over a 
larger span of spins in the retrograde case than in the prograde case.  This is consistent with the results presented in Ref.~\cite{SSTT,TSTS} where it was shown that the PN approximation waveforms perform worse in the retrograde case.  We should, therefore, not be surprised that the errors are greater for retrograde motion.
Now, referring once more to the equal mass case, the 
P-approximant templates outperform their T-approximant counterparts.
Even with this, we must again conclude that while on the surface there is no 
clear case for using P-approximant templates in searching for retrograde motion 
systems, they should be used because of the lower bias incurred in the estimation of 
parameters.

\subsection{Schwarzschild Orbits.}
While we have presented the maximized fitting factors and percentage errors in 
the estimation of the chirp-mass for the Schwarzschild case, we have not made 
any remarks regarding the maximized spin value.  In 
Table~\ref{table:schwarzoverlaps} we present a comparison of the maximized 
spin for both T and P-approximant templates for a selection of systems.  We can 
see from the table that the best overlap is never achieved using the 
Schwarzschild value for the spin.  For T-approximants the best fitting factors 
are all obtained by using retrograde values for the spin.  For P-approximants, 
it varies between prograde and retrograde values.  In general, regardless of 
whether the maximized spin is a prograde or retrograde value, we can see that 
the absolute error in the spin value is always smaller for P-approximant 
templates.

\begin{table}
\caption{Fitting factors for a variety of Schwarzschild systems of various 
masses.  The maximized values of $m_{1}$, $m_{2}$ and $q$ are presented in 
brackets underneath.} 
\begin{tabular}{c| c  c c c c }
\hline
\hline
$(m_{1},\, m_{2})$ & $\left< h_{T_{8}}\left| h^{\rm X}\right>\right.$ & 
$\left< h_{P_{8}}\left| h^{\rm X}\right>\right.$ &  $(m_1^T,\, m_2^T,\, q^T)$ & $(m_1^P,\, m_2^P,\, q^P)$ \\ \hline
(1.4,\, 50) &  0.998  &  0.999  & $ (1.5,\, 49.0,\, -0.02)$    & $ (1.4,\, 52.7,\, +0.06)$     \\
(1.4,\, 45) &  0.998  &  0.999  & $ (1.4,\, 48.6,\, -0.08)$    & $ (1.4,\, 46.1,\, +0.02)$     \\ 
(1.4,\, 40) &  0.997  &  0.999  & $ (1.5,\, 37.3,\, -0.09)$    & $ (1.4,\, 41.5,\, +0.04)$     \\
(1.4,\, 35) &  0.995  &  0.998  & $ (1.5,\, 32.8,\, -0.08)$    & $ (1.4,\, 36.0,\, +0.03)$     \\
(1.4,\, 30) &  0.978  &  0.996  & $ (1.6,\, 25.9,\, -0.15)$    & $ (1.4,\, 30.8,\, +0.03)$     \\ 
(1.4,\, 25) &  0.991  &  0.996  & $ (1.5,\, 22.6,\, -0.14)$    & $ (1.5,\, 24.9,\, -0.10)$     \\ 
(1.4,\, 20) &  0.990  &  0.995  & $ (1.5,\, 17.9,\, -0.16)$    & $ (1.4,\, 20.1,\, -0.01)$     \\ 
(1.4,\, 15) &  0.996  &  0.997  & $ (1.5,\, 13.5,\, -0.14)$    & $ (1.4,\, 15.0,\, -0.02)$     \\ 
(1.4,\, 10) &  0.996  &  0.999  & $ (1.6,\, 8.30,\, -0.19)$    & $ (1.4,\, 10.0,\, -0.02)$     \\
\hline
\hline
\end{tabular}
\label{table:schwarzoverlaps}
\end{table}


\section{Conclusions}\label{sec:conclusions}
We have applied P-approximant templates to the case of a NS orbiting Kerr BHs 
of varying mass and spin.  When graphically matching the numerical fluxes we 
found that the P-approximant fluxes gave a better fit as 
compared to the PN fluxes, especially when we varied the value of the spin 
parameter and best fitted the exact flux.  However, the true test of how 
well a template performs is reflected by the fitting factors achieved.  Using a 
signal waveform constructed from the exact expression for the orbital energy 
and numerical fluxes from black hole perturbation theory, we were able to 
compare the performance of T and P-approximant templates.  
In the case of retrograde, Schwarzschild and 
prograde orbits, not only did the P-approximants gave better and more reliable 
fitting factors, they also gave smaller biases in the estimation of parameters.  
We also saw the true power of the P-approximant templates in that we 
were able to generate templates right up to the LSO .  This is something 
that was not possible with the T-approximant templates due to the approximation 
for the flux function becoming negative before the LSO is reached.  
This restricted just how much of the numerical waveform we could model.  While not 
being completely correct due to the fact that we omitted the finite-mass correction 
terms, we also saw that the P-approximant templates gave the best 
performance when in the equal-mass case.  

The bias in the estimation of parameters is particularly poor in the case of
T-approximants especially for high-mass systems. Since high-mass systems can be
seen to a greater distance, and therefore likely to be the first sources
to be detected, it is important to keep in mind that the bias in the estimation
of parameters could be as large as the random errors associated with the measurement.
Such biases could seriously affect the test of general relativity \cite{BS95} that are
envisaged to be carried out with the aid of high-mass systems.
It is clear that for the type of systems examined in this paper, namely
equatorial test-mass circular orbits in Kerr, P-approximant templates are to be
preferred over their PN counterparts in both detection and measurement. It remains
to be seen if P-approximants (or their improved versions) prove to be good enough
when considering generic orbits in Kerr and for comparable mass systems with
spinning black holes on generic orbits.

\section*{Acknowledgments}
We would like to thank Prof. M. Shibata for his permission to use the numerical 
fluxes in this work.  We would also like to thank T. Damour, B. Iyer and the members of the Cardiff 
relativity group for the stimulating conversations and helpful suggestions.  
EKP would like to thank the School of Physics and Astronomy, Cardiff, for its 
hospitality in the final stages of this work. This research was supported in parts
by the Particle Physics and Astronomy Research Council and the Leverhulme Trust, 
both in the UK.


\appendix
\section{Pad\'e Approximation}
\label{sec:padeapp}
Let $f(x)$ be an analytic function whose Taylor expansion in $x,$
denoted $T_n(x),$ is $T_n(x) = \sum_{k=0}^n a_n x^n.$ 

We construct the Pad\'e Approximation $P^{N}_{M}(x)$ of $T_{n}(x)$ 
by equating the polynomial expansion with a 
rational function to the same order according to 
\begin{equation}
T_{n}(x) = P^{N}_{M}(x) \equiv \frac{B_{N}(x)}{D_{M}(x)},
\label{eq:ratfunc1}
\end{equation}
where
\begin{equation}
\newcommand\Dfrac[2]{\frac{\displaystyle #1}{\displaystyle #2}}
\frac{B_{N}(x)}{D_{M}(x)} = \Dfrac{\sum_{n=0}^{N} b_{n}\,x^{n}}{\\ 
\sum_{n=0}^{M} d_{n}\,x^{n}},
\label{eq:ratfunc2}
\end{equation}
such that $n+1 = N + M + 1$.  The benefit of using Pad\'e Approximants over 
Taylor Approximants is that the error in the approximation after $L$ terms 
reduces by a factor of $2^{L}$~\cite{BenOrz}.  If we let $M = N + \epsilon$, 
where $\epsilon = 0$ or $1$, we construct the ``diagonal'' and ``sub-diagonal'' 
Pad\'e Approximants respectively.  The  advantage of this is that we can 
cast the rational function in the form of a continued fraction, i.e.
\begin{equation}
\newcommand{\D}{\displaystyle}
P^{N}_{N + \epsilon} = \frac{c_{0}}{\D1 + 
                        \frac{c_{1}x}{\D1 + 
                         \frac{c_{2}x}{\D
                          \frac{\ddots}{{\D 1 + 
                                c_{n}x}}}}}.
\label{eq:confrac}
\end{equation}
If we use the rational function interpretation, Eq.~\ref{eq:ratfunc2}, each time 
we go to a higher order $n$ of approximation, we have to recalculate each of 
the coefficients, $b_{n}$ and $d_{n}$.  However, by using the continued 
fraction formalism, we only have to calculate the newest coefficient, $c_{n}$.  
As an example,  consider the Taylor expansion
\begin{equation}
T_{n}(v) = a_{0} + a_{1}(v).
\end{equation}
By matching this to a Pad\'e Approximation of the same order we get
\begin{equation}
a_{0} + a_{1}\,v = \frac{c_{0}}{1 + c_{1}v}.
\end{equation}
Solving for the Pad\'e coefficients, $c_{n}$, gives us
\begin{equation}
c_{0} = a_{0}\,\,\,\,\,\, , \,\,\,\,\,\,c_{1} = -\frac{a_1}{a_0}.
\end{equation}
If we now go to the next order, i.e.
\begin{equation}
\newcommand{\D}{\displaystyle}
a_{0} + a_{1}\,v + a_{2}\,v^{2} = \frac{c_{0}}{\D1 + \frac{c_{1}v}{\D1 + 
c_{2}v}},
\end{equation}
and solve for the coefficients $c_{n}$ once more, we obtain
\begin{equation}
c_{0} = a_{0}\,\,\,\,\,\, , \,\,\,\,\,\,c_{1} = 
-\frac{a_{1}}{a_{0}}\,\,\,\,\,\, , \,\,\,\,\,\,c_{2} = 
-\frac{a_{2}}{a_{1}}+\frac{a_{1}}{a_{0}}.
\end{equation}
The example shows that as we go to higher and higher orders in 
$v$, we only have to calculate one new coefficient as all the lower order 
coefficients remain constant. Consequently, there is a sense of
stability in the behaviour of the continued fraction coefficients
and convergence is easier to test.

There are, unfortunately, problems with Pad\'e approximation that
one should bear in mind:  Firstly, 
there are regions where Pad\'e approximants diverge while Taylor Approximants 
converge.  The second, and perhaps most important feature is that when
Pad\'e approximation does indeed converge, we cannot be sure that it 
converges to the exact function that we had in mind~\cite{NumRec}.  It should 
therefore always be kept in mind that there are inherent difficulties in using 
Pad\'e Approximants. Finally, Pad\'e approximants sometimes exhibit spurious
poles in the region of interest. However, as one will be able to test {\it a priori}
the existence of poles one can avoid the use of such approximants.
\end{document}